\documentclass[12pt,a4paper]{article}

\usepackage{authblk}
\usepackage{amsmath,amssymb}
\usepackage{graphicx}
\usepackage{amsfonts}
\usepackage{epsfig}

\usepackage{tikz}
\usepackage{pgflibrarysnakes}
\usetikzlibrary[snakes]

\usetikzlibrary{arrows,shapes,positioning}
\usetikzlibrary{decorations.markings}
\tikzstyle arrowstyle=[scale=1]
\tikzstyle directed=[postaction={decorate,decoration={markings,
		mark=at position .65 with {\arrow[arrowstyle]{stealth}}}}]
\tikzstyle reverse directed=[postaction={decorate,decoration={markings,
		mark=at position .65 with {\arrowreversed[arrowstyle]{stealth};}}}]

\renewcommand{\baselinestretch}{1.5}

\newcommand{\tr}{\operatorname{Tr}}

\newcommand{\ket}[1]{\left|#1\right\rangle}      
\newcommand{\bra}[1]{\left\langle #1\right|}     
\newcommand{\eq}{\begin{equation}}
	\newcommand{\en}{\end{equation}}
\newcommand{\bear}{\begin{eqnarray}}
	\newcommand{\ear}{\end{eqnarray}}

\title{\mbox{}On the partition function of the $Sp(2n)$ integrable vertex model} 

\author{G.A.P. Ribeiro\footnote{E-mail: pavan@df.ufscar.br}}
\affil{\small Departamento de F\'isica, Universidade Federal de S\~ao Carlos, PO Box 676, 13565-905, S\~ao Carlos-SP, Brazil\vspace{8pt}}

\date{}

\begin{document}
	\renewcommand{\baselinestretch}{1.5}
	\maketitle
	
	\vspace{-.5in}\noindent
\begin{abstract}
We study the partition function per site of the integrable $Sp(2n)$ vertex model on the square lattice. We establish a set of transfer matrix fusion relations for this model. The solution of these functional relations in the thermodynamic limit allows us to compute the partition function per site of the fundamental $Sp(2n)$ representation of the vertex model. In addition, we also obtain the partition function of vertex models mixing the fundamental with other representations.
\end{abstract}
	
	\centerline{Keywords: Integrability, vertex models, fusion, inversion relations }
	\renewcommand{\baselinestretch}{1.5}
	
	\thispagestyle{empty}
	
	\newpage
	
	\pagestyle{plain}
	
	\pagenumbering{arabic}

	\section{Introduction}

The use of transfer matrix techniques to study statistical models on the square lattice has a long history \cite{BAXTER,BOOK-KBI}. This includes the use Bethe ansatz techniques, $T$-$Q$ relations and functional approaches. The latter have been shown to be successful in solving models with subtle algebraic and analytical structure. In this context, there are approaches based on inversion relations~\cite{STROGANOV,BAXTER-inversion,SHANKAR,KLUMPER1990}, fusion functional equations~\cite{BAXTER-PEARCE,KIRILLOV,BAZHANOV} and transfer matrix inversion identities~\cite{PEARCE1987}.

Recently it was shown \cite{RKP} that another set of relations, the transfer matrix fusion identities, can be used to study integrable vertex models in the thermodynamic limit with subtle analyticity properties.
It was noticed in \cite{RKP} that the leading eigenvalue of the fundamental representation of the integrable $Sp(4)$ vertex model displays an extended singularity at the center of the analyticity strip, which prevents the use of the usual inversion relation to solve the problem. The existence of the extended singularity splits the analyticity strip into two parts, which requires extra relations to connect both sides of the analyticity strip. The transfer matrix fusion identities in the thermodynamic limit, which is an exact truncation of the fusion hierarchy \cite{KAROWSKI,KULISH1981,KULISH1982,CAO2019}, constitutes a set of sufficient relations that allowed for the computation of the partition function per site in the thermodynamic limit of the $Sp(4)$ vertex model on the square lattice \cite{RKP}. It is remarkable that the obtained solution exhibits explicitly a kind of CDD factor due to the loss of analyticity along an infinitely long line at the center of the analytical strip.

In this work, we are interested in tackling the general $Sp(2n)$ integrable vertex model, which therefore generalizes the results obtained in \cite{RKP}. In order to do that, we first derive the transfer matrix fusion relations for the integrable $Sp(6)$ vertex model. These relations hold for arbitrary values of the spectral parameter, which is in contrast with the discrete set of relations used in \cite{CAO}. These fusion relations are naturally extended to the arbitrary $Sp(2n)$ vertex model. In the thermodynamic limit, the transfer matrix fusion relations become an exact truncation of the fusion hierarchy. Remarkably, these relations are also just enough to allow for the computation of the partition function per site of the $Sp(2n)$ vertex model. Apart from the solution for the last fusion level, the solution for all other fusion levels shows a kind of CDD factor due to the loss of analyticity along an extended singularity at the center of the strip. This is due to the fact that only the eigenvalue of the fusion transfer matrix of the last fusion level is free of zeros inside the analyticity strip. This is described in detail in the case of the $Sp(6)$ vertex model and the general formulae are given for the arbitrary $Sp(2n)$ case.
 
This paper is organized as follows. In section \ref{INTEGRA}, we described the integrable structure of the model. In section \ref{sp6} we deal with the $Sp(6)$ case.  We discuss the fusion properties and the transfer matrix fusion identities for the $Sp(6)$ model. We study the analyticity of the leading eigenvalues of the fundamental and fused transfer matrices. The partition function per site is evaluated in the thermodynamic limit. In section \ref{general} we extend the results to the arbitrary $Sp(2n)$ vertex model. Our conclusions are given in section \ref{CONCLUSION}. Additional details are given in the appendices.

\section{The vertex model}\label{INTEGRA}

The fundamental $Sp(2n)$ integrable vertex model is described by the $R$-matrix \cite{RESHETIKHIN,KUNIBA,KULISH,MARTINS1997},
\eq
R_{12}^{(2n,2n)}(\lambda)= \lambda (\lambda+\Delta) I_{12} + (\lambda+\Delta) P_{12} +\lambda E_{12},
\label{Rmatrix}
\en
which acts in the indicated spaces of the tensor product $W\otimes W$, 
where $W$ is the fundamental representation of the $Sp(2n)$, which is  of dimension $2n$ as indicated in the superscript. The parameter $\Delta=n+1$  and $I_{i,i+1}$, $P_{i,i+1}$ and $E_{i,i+1}$ are the identity, permutation and Temperley-Lieb operators acting on the sites $i$ and $i+1$.  Their matrix elements are given as $(I_{i,i+1})_{ac}^{bd}=\delta_{a,b}\delta_{c,d}$, $(P_{i,i+1})_{ac}^{bd}=\delta_{a,d}\delta_{b,c}$ and $(E_{i,i+1})_{ac}^{bd}=\epsilon_{a} \epsilon_c \delta_{a,2n+1-c}\delta_{b,2n+1-d}$ for $1\leq a,b,c,d \leq n$ where $\epsilon_a=1$ for $1\leq a \leq n$ and $\epsilon_a=-1$ for $n+1\leq a \leq 2n$. 

The $R$-matrix has the important properties of regularity, unitarity and crossing given as follows,
\bear
R_{12}^{(2n,2n)}(0)&=& \Delta P_{12}, \\
R_{12}^{(2n,2n)} (\lambda) R_{21}^{(2n,2n)} (-\lambda) &=& (1-\lambda^2) (\Delta^2-\lambda^2)I_{12}, \\
R_{12}^{(2n,2n)} (\lambda) &=& (V\otimes I) (R^{(2n,2n)}_{12}(-\lambda-\rho) )^{t_2} (V^{-1}\otimes I),
\ear
where $t_2$ is transposition in the second space, the crossing parameter is $\rho= \Delta$ and the crossing matrix $V$ is given by $V=\mbox{anti-diagonal}(1,\dots,1,-1,\dots,-1)$, where the matrix entries are listed from the top-right to the bottom-left corners. The $R$-matrix satisfies the Yang-Baxter equation,
\eq
R^{(2n,2n)}_{12}(\lambda-\mu) R^{(2n,2n)}_{13} (\lambda) R^{(2n,2n)}_{23} (\mu) =R^{(2n,2n)}_{23}(\mu) R^{(2n,2n)}_{13}(\lambda)  R^{(2n,2n)}_{12} (\lambda-\mu).
\label{yang-baxter}
\en

The partition function of the classical $M\times L$ lattice model with periodic boundary conditions in both directions can be written as $Z=\tr{\left[\left(T^{(2n)}(\lambda)\right)^M\right]}$, where $T^{(2n)}(\lambda)$ is the row-to-row transfer matrix given by the trace over the $2n$-dimensional auxiliary space $\cal A$  of the monodromy matrix  ${\cal T}_{\cal A}^{(2n,2n)}(\lambda)=R_{{\cal A}L}^{(2n,2n)}(\lambda)\dots R_{{\cal A} 1}^{(2n,2n)}(\lambda)$ such as,
\eq
T^{(2n)}(\lambda)=\tr_{\cal A}{[ {\cal T}_{\cal A}^{(2n,2n)}(\lambda)]}.
\en
The transfer matrix constitutes a family of commuting operators $[T^{(2n)}(\lambda),T^{(2n)}(\mu)]=0$ thanks to the Yang-Baxter equation. Therefore, $T^{(2n)}(\lambda)$ is a generating function of conserved charges. The first non-trivial conserved charge is obtained by logarithmic derivative transfer matrix, ${\cal H}^{(2n)}=\frac{d}{d\lambda}\log{T^{(2n)}(\lambda)}\Big|_{\lambda=0}$, which is the Hamiltonian  of the integrable $Sp(2n)$ spin chain with periodic boundary condition \cite{RESHETIKHIN,KUNIBA,KULISH,MARTINS1997,MARTINS,MARTINS-SP2N},
\eq
{\cal H}^{(2n)}=\sum_{i=1}^{L}\left(\frac{1}{\Delta}I_{i,i+1} +P_{i,i+1} - \frac{1}{\Delta} E_{i,i+1}\right),
\en
whose physical properties were studied via the solution of the Bethe ansatz equation in \cite{MARTINS-SP2N}.

\section{$Sp(6)$ vertex model}\label{sp6}

\subsection{Fusion relations}\label{fusion}

For the $Sp(6)$ case, the tensor product of two fundamental representations decomposes as $6\otimes 6 = 1 \oplus 14 \oplus 21$ ~\cite{GROUP}. 
This means that we can rewrite the fundamental
$R$-matrix $R^{(6,6)}_{12}(\lambda)$ in terms of the projectors in such spaces, namely
\eq
R^{(6,6)}_{12}(\lambda)=(\lambda+1)(\lambda-4) \check{P}_{12}^{(1)} + (\lambda-1)(\lambda+4) \check{P}_{12}^{(14)}  + (\lambda+1)(\lambda+4) \check{P}_{12}^{(21)},  
\label{r66}
\en
where $\check{P}_{12}^{(\alpha)}$ are the projectors on the  $\alpha$-dimensional subspace ($\alpha=1,14,21$), which are given  in the Appendix~A. This shows explicitly the singular values that degenerate in projection operators, which implies $R^{(6,6)}_{12}(-4) =24 \check{P}_{12}^{(1)}$ and $R^{(6,6)}_{12}(-1) =-6 \check{P}_{12}^{(14)}$.

By the rules of fusion \cite{KAROWSKI,KULISH1981,KULISH1982}, one can exploit the point $\lambda=-1$ to obtain a new $R$-matrix with a $14$-dimensional auxiliary space (see \cite{CAO,ODBA} and the Appendix~A for more details on the fusion rules for $Sp(6)$) given as,
\eq
R_{12}^{(14,6)}(\lambda)=(\lambda+\frac{3}{2})(\lambda-\frac{7}{2}) \check{P}_{12}^{(6)} + (\lambda-\frac{3}{2})(\lambda+\frac{7}{2})\check{P}_{12}^{({14}')} +  (\lambda+\frac{3}{2})(\lambda+\frac{7}{2})\check{P}_{12}^{(64)} ,
\en
where the projectors $\check{P}_{12}^{(6)}$, $\check{P}_{12}^{(14')}$ and $\check{P}_{12}^{(64)}$  (also given in the Appendix~A) are due to the decomposition $14\otimes 6 = 6 \oplus 14' \oplus 64 $. It is interesting to note that there are two different $14$-dimensional irreducible representations. In terms of the Dynkin labels, the fundamental $6$-dimensional and the two $14$-dimensional representations are given by $(6)=[1,0,0]$, $(14)=[0,1,0]$ and $(14')=[0,0,1]$ respectively. 
One can also simply read the singular values, such that $R_{12}^{(14,6)}(-\frac{7}{2})=14 \check{P}_{12}^{(6)}$ and $R_{12}^{(14,6)}(-\frac{3}{2})=-6 \check{P}_{12}^{(14')}$.

Finally, due to the tensor product decomposition of $14'\otimes 6= 14 \oplus 70$, one has that the $R$-matrix $R_{12}^{(14',6)}(\lambda)$ is given in terms of the respective projectors as follows,
\eq
R_{12}^{(14',6)}(\lambda) = (\lambda-3) \check{P'}_{12}^{(14)} + (\lambda+3) \check{P}_{12}^{(70)}.
\en
Here it is important to note, that $\check{P'}_{12}^{(14)}$ is the projector from the $14\times 6$-dimensional space back to the $(14)=[0,1,0]$ representation, while $\check{P}_{12}^{(14)}$ in (\ref{r66}) is the projector from the $6\times6$-dimensional space to the $(14)=[0,1,0]$ representation.

Naturally, these fused $R$-matrices also satisfy the unitarity conditions, 
\bear
R_{12}^{(14,6)} (\lambda) R_{21}^{(6,14)} (-\lambda) &=& ((\tfrac{3}{2})^2-\lambda^2)((\tfrac{7}{2})^2-\lambda^2) I_{12}, \\
R_{12}^{(14',6)} (\lambda) R_{21}^{(6,14')} (-\lambda) &=& ((3)^2-\lambda^2) I_{12},
\ear
and Yang-Baxter equations,
\eq
R_{12}^{(\alpha,6)}(\lambda-\mu) R_{13}^{(\alpha,6)} (\lambda) R_{23}^{(6,6)} (\mu) =R_{23}^{(6,6)}(\mu) R_{13}^{(\alpha,6)}(\lambda)  R_{12}^{(\alpha,6)} (\lambda-\mu),
\label{yang-baxter2}
\en
for $\alpha=14,14'$. This allows us to define two additional transfer matrices with $14$-dimensional auxiliary spaces.
\eq
T^{(\alpha)}(\lambda)=\tr_{\cal A}{\left[{\cal T}_{\cal A}^{(\alpha,6)}(\lambda)\right]}, \qquad {\cal T}_{\cal A}^{(\alpha,6)}(\lambda)=R_{{\cal A}L}^{(\alpha,6)}(\lambda)R_{{\cal A}L-1}^{(\alpha,6)}(\lambda)\cdots R_{{\cal A}1}^{(\alpha,6)}(\lambda) .         
\en
The above transfer matrices $T^{(6)}(\lambda)$, $T^{(14)}(\lambda)$ and $T^{(14')}(\lambda)$ also commute mutually for different spectral parameters.

The fusion structure allows us to establish the  transfer matrix fusion identities for $Sp(6)$ along the same lines as \cite{RKP}, which are given by
\bear
T^{(6)}(\lambda)T^{(6)}(\lambda-4)&=& [(\lambda^2-1) (\lambda^2-4^2)]^L I(1 + O(e^{-L})), \label{TMFISp61} \\
T^{(6)}(\lambda)T^{(6)}(\lambda-1)&=& [(\lambda-1) (\lambda+4)]^L T^{(14)}(\lambda-\frac{1}{2})(1+O(e^{-L})), \label{TMFISp62}\\
T^{(6)}(\lambda)T^{(14)}(\lambda-\frac{3}{2})&=& [(\lambda^2-1) (\lambda+4)]^L T^{(14')}(\lambda-1) (1+O(e^{-L})), \label{TMFISp63}\\
T^{(6)}(\lambda)T^{(14)}(\lambda-\frac{7}{2})&=& [(\lambda-1)(\lambda+4)]^L T^{(6)}(\lambda-3) (1+O(e^{-L})), \label{TMFISp64} \\
T^{(6)}(\lambda)T^{(14')}(\lambda-3)&=& [(\lambda+4)]^L T^{(14)}(\lambda-\frac{5}{2}) (1+O(e^{-L})), \label{TMFISp65}
\ear
where (\ref{TMFISp61}) is the transfer matrix inversion identity \cite{PEARCE1987} and $O(e^{-L})$ are corrections vanishing exponentially in the thermodynamic limit. The additional relations (\ref{TMFISp62}-\ref{TMFISp65}) are derived along the same lines as \cite{RKP} by exploiting the singular values 
$\lambda=-1$ of the $R$-matrix $R_{12}^{(6,6)}(\lambda)$, $\lambda=-\frac{3}{2}$ and $\lambda=-\frac{7}{2}$ of $R_{12}^{(14,6)}(\lambda)$ and $\lambda=-3$ of $R_{12}^{(14',6)}(\lambda)$ (see Appendix~A).

As a consequence of the commutativity property of the transfer matrices $T^{(6)}(\lambda)$, $T^{(14)}(\lambda)$ and $T^{(14')}(\lambda)$, their common eigenvectors are independent of the spectral parameter $\lambda$. This implies that the relations (\ref{TMFISp61}-\ref{TMFISp65}) are also satisfied by the transfer matrix eigenvalues $\Lambda^{(\alpha)}(\lambda)$, 
\bear
\Lambda^{(6)}(\lambda)\Lambda^{(6)}(\lambda-4)&=& [(\lambda^2-1) (\lambda^2-4^2)]^L \left(1 + O(e^{-L})\right), \label{TMFRSp61} \\
\Lambda^{(6)}(\lambda)\Lambda^{(6)}(\lambda-1)&=& [(\lambda-1) (\lambda+4)]^L \Lambda^{(14)}(\lambda-\frac{1}{2})\left(1 + O(e^{-L})\right), \label{TMFRSp62}\\
\Lambda^{(6)}(\lambda)\Lambda^{(14)}(\lambda-\frac{3}{2})&=& [(\lambda^2-1) (\lambda+4)]^L \Lambda^{(14')}(\lambda-1) \left(1 + O(e^{-L})\right), \label{TMFRSp63}\\
\Lambda^{(6)}(\lambda)\Lambda^{(14)}(\lambda-\frac{7}{2})&=& [(\lambda-1)(\lambda+4)]^L \Lambda^{(6)}(\lambda-3)\left(1 + O(e^{-L})\right), \label{TMFRSp64} \\
\Lambda^{(6)}(\lambda)\Lambda^{(14')}(\lambda-3)&=& [(\lambda+4)]^L \Lambda^{(14)}(\lambda-\frac{5}{2})\left(1 + O(e^{-L})\right), \label{TMFRSp65}
\ear
which hold for $|\lambda|<\epsilon$ for some small fixed $\epsilon$. Note that (\ref{TMFRSp61}) is the inversion relation \cite{STROGANOV,BAXTER-inversion,SHANKAR}
for the $Sp(6)$ vertex model. It is worth recalling that the above relations hold for all the transfer matrix eigenvalues for large $L$ and they are an exact truncation of the fusion hierarchy in the limit $L\rightarrow\infty$.

\begin{figure}[t]
	\begin{center}
		\begin{minipage}{0.33\linewidth}
			\includegraphics[width=0.74\linewidth, angle=-90]{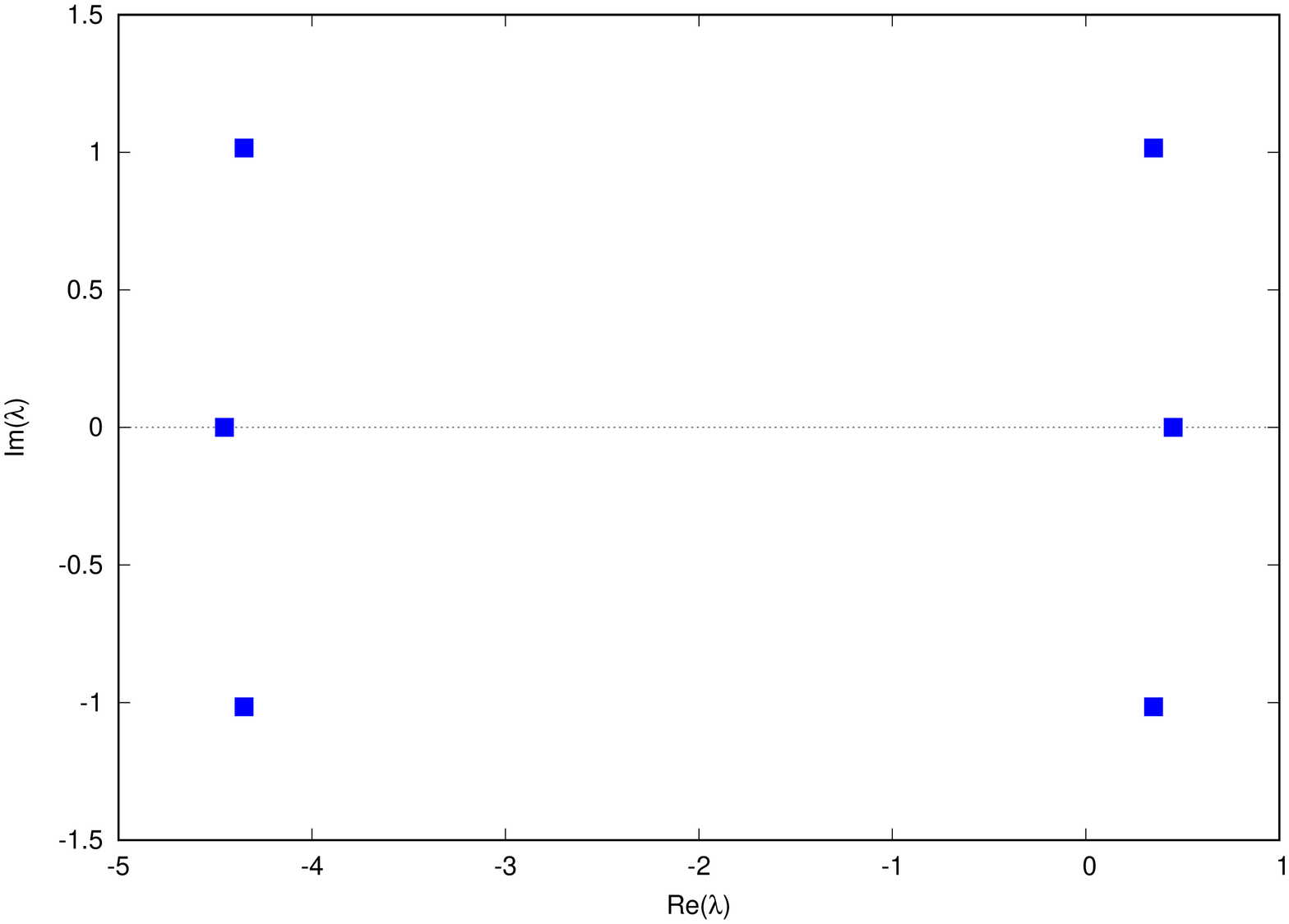}
			\includegraphics[width=0.74\linewidth, angle=-90]{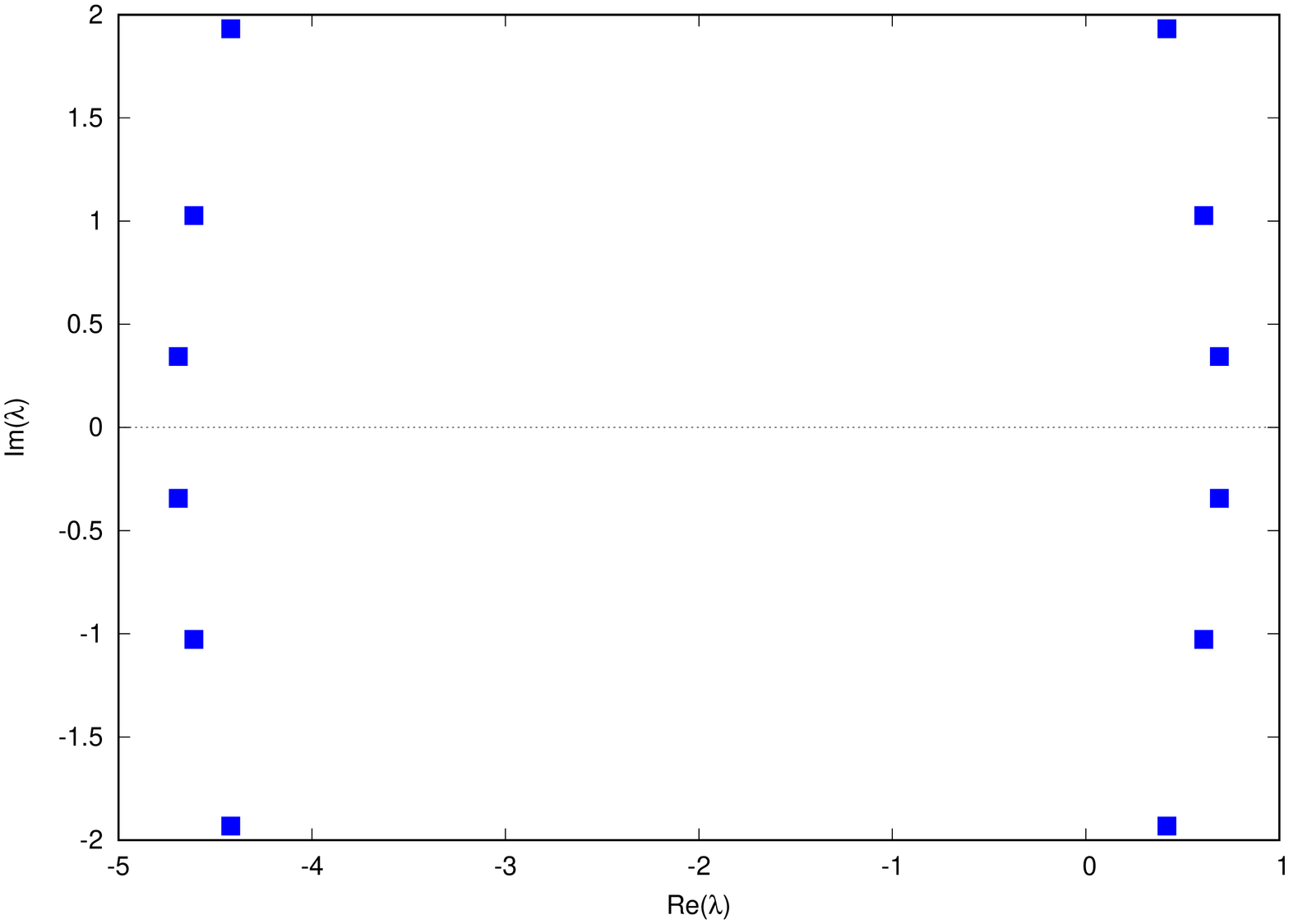}
		\end{minipage}%
		\begin{minipage}{0.33\linewidth}
			\includegraphics[width=0.74\linewidth, angle=-90]{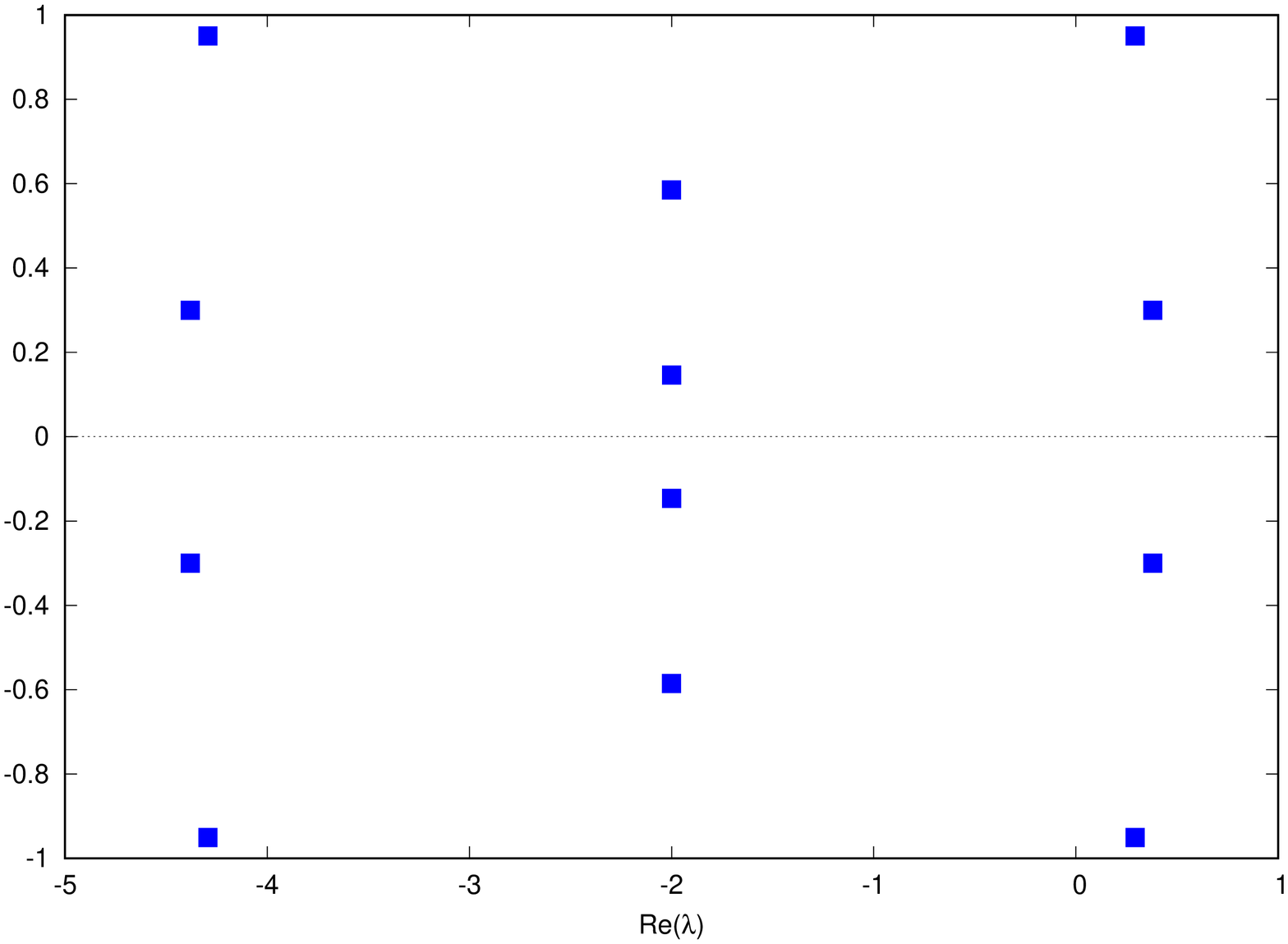}
			\includegraphics[width=0.74\linewidth, angle=-90]{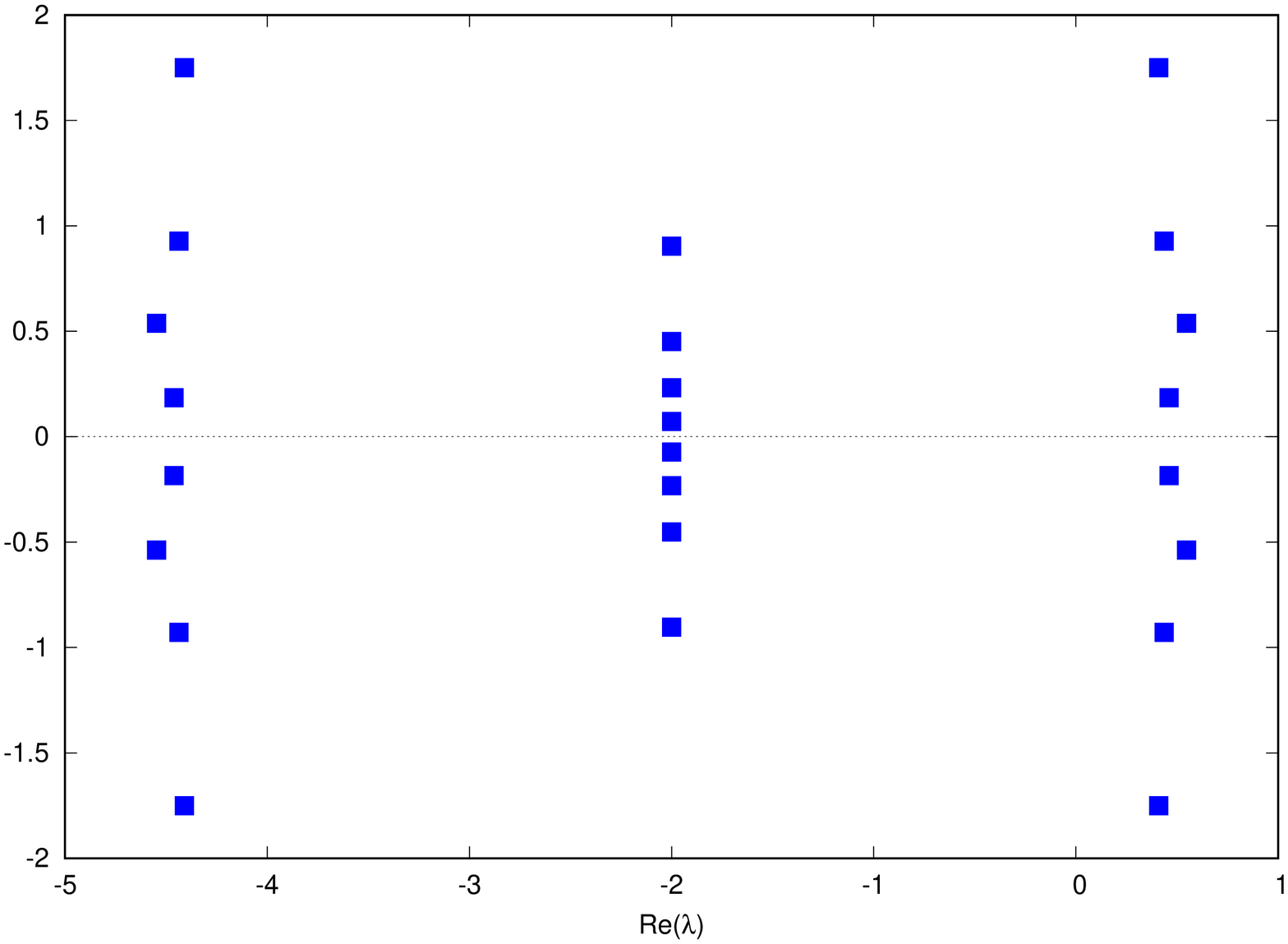}
		\end{minipage}%
		\begin{minipage}{0.33\linewidth}
			\includegraphics[width=0.74\linewidth, angle=-90]{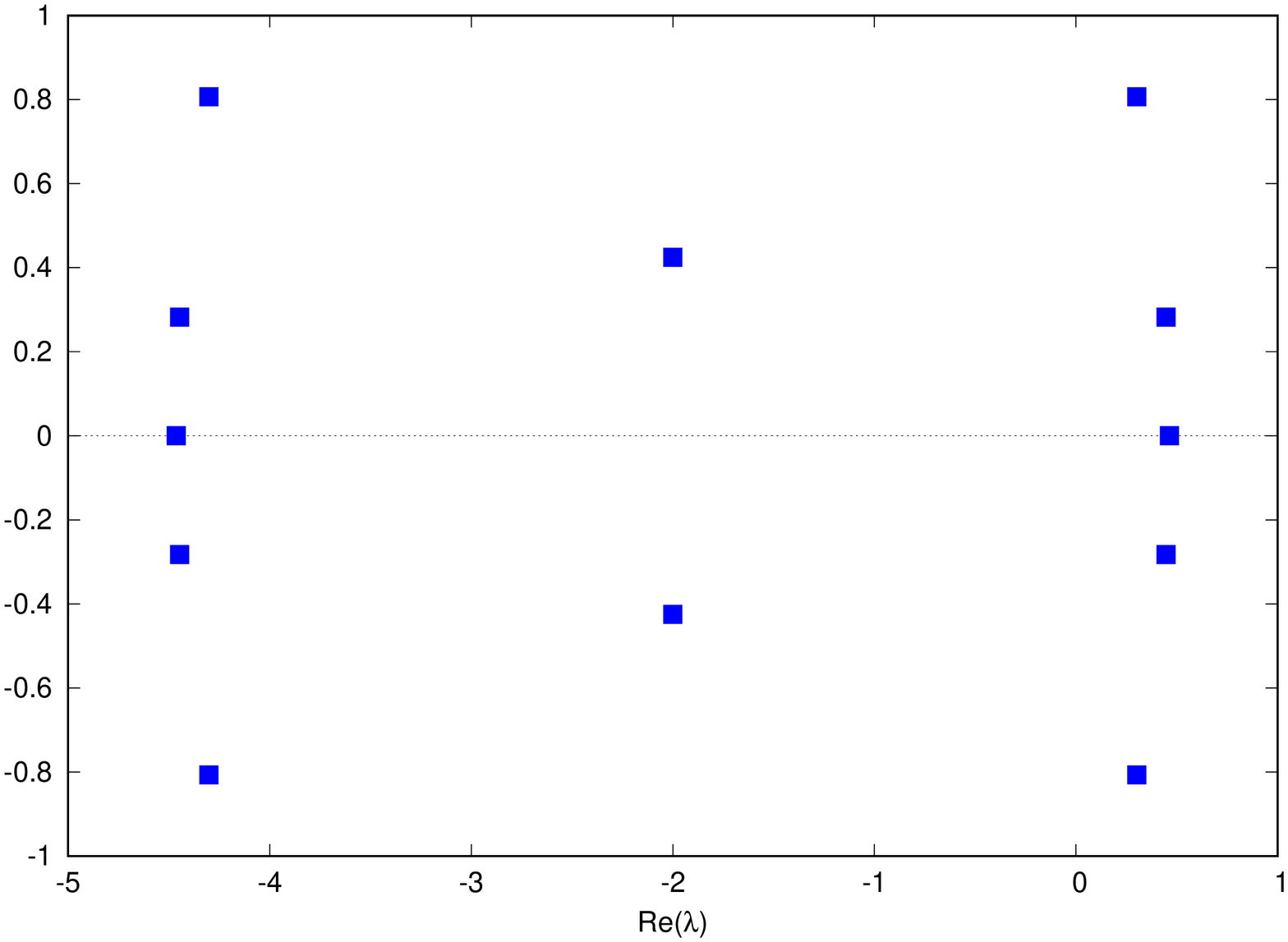}
			\includegraphics[width=0.74\linewidth, angle=-90]{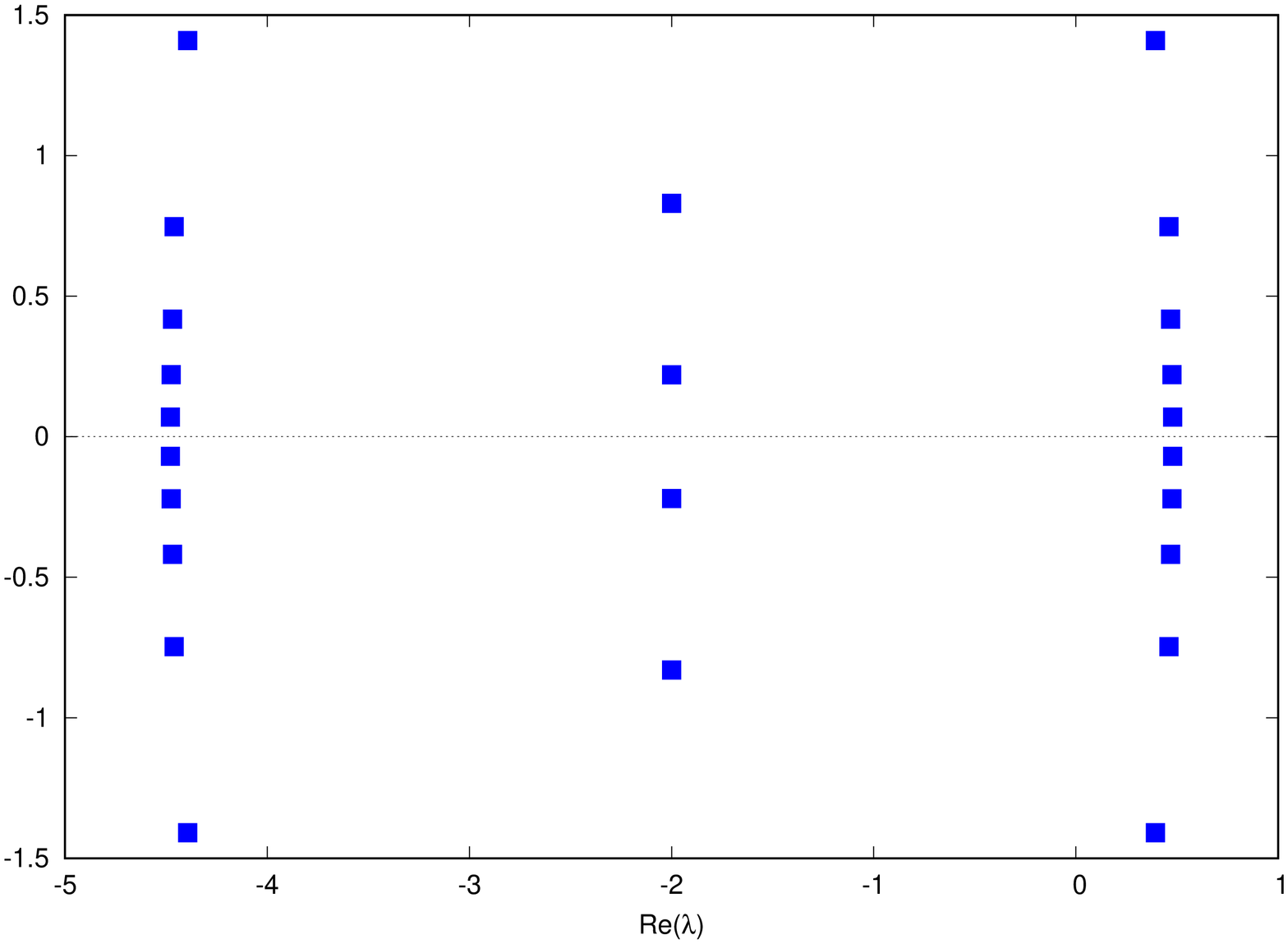}
		\end{minipage}
		\caption{Left panel: Zeros of the leading eigenvalue $\Lambda_0^{(14')}(\lambda)$ for $L=6,12$. Center panel: Zeros of the leading eigenvalue $\Lambda_0^{(14)}(\lambda)$ for $L=6,12$ exhibiting $2 L/3$ zeros in the center of the analytical strip.
			Right panel: Zeros of the leading eigenvalue $\Lambda_0^{(6)}(\lambda)$ in the complex plane for $L=6,12$ exhibiting $L/3$ zeros in the center of the analytical strip. }
		\label{fig1}
	\end{center}
\end{figure}

We studied the pattern of zeros of the largest eigenvalue of the transfer matrix for the $Sp(6)$ vertex model. Our results are exhibited in Figure~\ref{fig1} for lattice sizes
$L=6$ and $L=12$ respectively. This study reveals that, for the $Sp(6)$ vertex model, there are zeros of $\Lambda_0^{(6)}(\lambda)$ and $\Lambda_0^{(14)}(\lambda)$ precisely at the center of the analyticity strip ($-\frac{9}{2}<\mbox{Re}(\lambda)<\frac{1}{2}$) along the line $\Re(\lambda)=-2$ and the number of those zeros and their density grows linearly with system size. This accumulation of zeros breaks the analyticity along the center line of the analytical strip. This special behaviour was first realized for the $Sp(4)$ vertex model \cite{RKP} and it is also the case for $Sp(6)$. 
We note that only the leading eigenvalue $\Lambda_0^{(14')}(\lambda)$ of the $[0,0,1]$ representation is free of zeros inside the wider strip $-\frac{9}{2}<\mbox{Re}(\lambda)<\frac{1}{2}$, which implies that the set of relations (\ref{TMFRSp61}--\ref{TMFRSp65}) are just enough to determine the leading eigenvalues $\Lambda^{(6)}_0(\lambda)$, $\Lambda^{(14)}_0(\lambda)$ and $\Lambda_0^{(14')}(\lambda)$. 

\subsection{Partition function}\label{limit}

In this section we determine the largest eigenvalue $\Lambda_0^{(6)}(\lambda)$ of the transfer matrix in the extended region $-\frac{9}{2}<\mbox{Re}(\lambda)<\frac{1}{2}$ and as a byproduct we also obtain the largest eigenvalues $\Lambda_0^{(14)}(\lambda)$ and $\Lambda_0^{(14')}(\lambda)$.

In order to do this, we conveniently define the partition function per site and its logarithmic derivative
\eq
\kappa^{(\alpha)}(\lambda)= \lim_{L\rightarrow\infty} \left(\Lambda_0^{(\alpha)}(\lambda)\right)^{1/L}\!\!, \quad\; 
\omega^{(\alpha)}(\lambda)=\frac{d}{d\lambda}\log{\kappa^{(\alpha)}}(\lambda),\quad\; \mbox{for $\alpha=6,14$ and $14'$}. \label{kappaomega}
\en
Having in mind the break in analyticity, we further define
\eq
\kappa^{(\alpha)}(\lambda)=\begin{cases}
	\kappa^{(\alpha)}_I(\lambda),&\lambda<-2\\
	\kappa^{(\alpha)}_{II}(\lambda),&\lambda>-2
\end{cases},\qquad
\omega^{(\alpha)}(\lambda)=\begin{cases}
	\omega^{(\alpha)}_I(\lambda),&\lambda<-2\\
	\omega^{(\alpha)}_{II}(\lambda),&\lambda>-2
\end{cases},\label{break}
\en 
where the indices $I$ and $II$ specify the functions on the left and right of the cut line $\mbox{Re}(\lambda)=-2$ with $\omega_{i}^{(\alpha)}(\lambda)=\frac{d}{d\lambda}\log{\kappa_i^{(\alpha)}}(\lambda)$ for $i=I$ or $II$ and $\alpha=6,14$. 

Similarly to \cite{RKP}, the patterns of zeros of $\Lambda_0^{(\alpha)}(\lambda)$ for $\alpha=6,14,14'$ in Figure~\ref{fig1} are invariant under the crossing involution $\lambda\mapsto -4-\lambda$, which implies the following symmetries,
\begin{alignat}{3}
	&\hspace{-.1in}\kappa_{I}^{(\alpha)}(\lambda)=\kappa_{II}^{(\alpha)}(-4-\lambda), &&\kappa_{II}^{(\alpha)}(\lambda)=\kappa_{I}^{(\alpha)}(-4-\lambda), \label{symmetries1}\\
	&\mbox{}\hspace{-.1in}\omega_{I}^{(\alpha)}(\lambda)=-\omega_{II}^{(\alpha)}(-4-\lambda),\quad &&\omega_{II}^{(\alpha)}(\lambda)=-\omega_{I}^{(\alpha)}(-4-\lambda),\label{symmetries2}
\end{alignat}
for $\alpha=6,14$ and $\kappa^{(14')}(\lambda)=\kappa^{(14')}(-4-\lambda)$  and $\omega^{(14')}(\lambda)=-\omega^{(14')}(-4-\lambda)$.

Using these functions, the fusion relations (\ref{TMFRSp61}-\ref{TMFRSp65}) can be rewritten  as
\bear
\kappa_{II}^{(6)}(\lambda)\kappa_{I}^{(6)}(\lambda-4)&=& (\lambda^2-1) (\lambda^2-4^2) , \label{psiMFISp61} \\
\kappa_{II}^{(6)}(\lambda)\kappa_{II}^{(6)}(\lambda-1)&=& (\lambda-1) (\lambda+4) \kappa_{II}^{(14)}(\lambda-\frac{1}{2}), \label{psiMFISp62}\\
\kappa_{II}^{(6)}(\lambda)\kappa_{II}^{(14)}(\lambda-\frac{3}{2})&=& (\lambda^2-1) (\lambda+4) \kappa^{(14')}(\lambda-1), \label{psiMFISp63}\\
\kappa_{II}^{(6)}(\lambda)\kappa_{I}^{(14)}(\lambda-\frac{7}{2})&=& (\lambda-1)(\lambda+4) \kappa_{I}^{(6)}(\lambda-3), \label{psiMFISp64} \\
\kappa_{II}^{(6)}(\lambda)\kappa^{(14')}(\lambda-3)&=& (\lambda+4) \kappa_{I}^{(14)}(\lambda-\frac{5}{2}), \label{psiMFISp65}
\ear
and its logarithmic derivative,
\bear
\omega_{II}^{(6)}(\lambda)+\omega_{I}^{(6)}(\lambda-4)&=&\frac{1}{\lambda+1}+\frac{1}{\lambda-1}+ \frac{1}{\lambda+4}+\frac{1}{\lambda-4}, \label{omegaMFISp61} \\
\omega_{II}^{(6)}(\lambda)+\omega_{II}^{(6)}(\lambda-1)&=& \frac{1}{\lambda-1}+\frac{1}{\lambda+4}+ \omega_{II}^{(14)}(\lambda-\frac{1}{2}), \label{omegaMFISp62}\\
\omega_{II}^{(6)}(\lambda)+\omega_{II}^{(14)}(\lambda-\frac{3}{2})&=& \frac{1}{\lambda+1}+\frac{1}{\lambda-1}+ \frac{1}{\lambda+4}+  \omega^{(14')}(\lambda-1), \label{omegaMFISp63}\\
\omega_{II}^{(6)}(\lambda)+\omega_{I}^{(14)}(\lambda-\frac{7}{2})&=& \frac{1}{\lambda-1}+\frac{1}{\lambda+4}+ \omega_{I}^{(6)}(\lambda-3), \label{omegaMFISp64} \\
\omega_{II}^{(6)}(\lambda)+\omega^{(14')}(\lambda-3)&=& \frac{1}{\lambda+4}+ \omega_{I}^{(14)}(\lambda-\frac{5}{2}). \label{omegaMFISp65}
\ear

By elimination, one can obtain a simpler equation for the $\omega^{(14')}(\lambda)$,
\eq
\omega^{(14')}(\lambda-1)+\omega^{(14')}(\lambda-5)=\frac{1}{\lambda+2}+\frac{1}{\lambda-4}.\label{omega14pFuncEq}
\en

The logarithmic derivatives have the advantage that they admit Fourier-Laplace transforms. Therefore, we Fourier-Laplace transform the above equations (\ref{omegaMFISp61}-\ref{omegaMFISp65}). The system of the resulting equations is algebraically
resolved. The algebraic expressions are finally transformed back and written in terms of gamma functions or integrals (as described in detail in \cite{RKP}). The final expression for $\omega_{II}^{(6)}(\lambda)$ written in terms of the gamma functions is given as follows,
\bear
\omega_{II}^{(6)}(\lambda)&=&\frac{d}{d\lambda}\log\left[\frac{ \Gamma(\frac{1}{3}+\frac{\lambda}{3})\Gamma(\frac{2}{3}-\frac{\lambda}{3})}{\Gamma(\frac{1}{3}-\frac{\lambda}{3}) \Gamma(\frac{2}{3}+\frac{\lambda}{3})}\right] \nonumber \\
&+& \frac{d}{d\lambda}\log\left[\frac{ \Gamma(\frac{7}{8}+\frac{\lambda}{8})\Gamma(\frac{1}{8}-\frac{\lambda}{8}) \Gamma(\frac{1}{4}+\frac{\lambda}{8}) \Gamma(\frac{3}{4} -\frac{\lambda}{8})  \Gamma(\frac{5}{8}+\frac{\lambda}{8}) \Gamma(\frac{3}{8}-\frac{\lambda}{8})}
{  \Gamma(\frac{7}{8}-\frac{\lambda}{8}) \Gamma(\frac{1}{8}+\frac{\lambda}{8}) \Gamma(\frac{1}{4}-\frac{\lambda}{8}) \Gamma(\frac{3}{4} +\frac{\lambda}{8}) \Gamma(\frac{5}{8}-\frac{\lambda}{8}) \Gamma(\frac{3}{8}+\frac{\lambda}{8}) }\right] \label{solsp6}\\
&+& \frac{d}{d\lambda}\log\left[\frac{ \Gamma(\frac{9}{8}+\frac{\lambda}{8}) \Gamma(\frac{5}{8}-\frac{\lambda}{8}) \Gamma(\frac{3}{2} + \frac{\lambda}{8}) \Gamma(1-\frac{\lambda}{8})}{ \Gamma(\frac{1}{8}-\frac{\lambda}{8}) \Gamma(\frac{5}{8}+\frac{\lambda}{8}) \Gamma(\frac{1}{2} - \frac{\lambda}{8}) \Gamma(1+\frac{\lambda}{8})} \right]. \nonumber
\ear
It is worth noticing that the last term in (\ref{solsp6}) would be the solution of the Eq.(\ref{omegaMFISp61}) if the eigenvalues were analytical in the entire strip. Therefore, the other terms in (\ref{solsp6}) can be seen as CDD factors \cite{CDD} due to the break of the analyticity properties at the branch cut line at center of the analyticity strip $-\frac{9}{2}<\mbox{Re}(\lambda)<\frac{1}{2}$.   

The ground state energy of the associated quantum spin chain is obtained by setting $\lambda=0$,
\bear
\omega_{II}^{(6)}(0)=\frac{5}{4}-\frac{\pi}{4}+ \frac{\pi}{2\sqrt2} -\frac{2 \pi}{3\sqrt3}-\frac{\log2}{2}-\frac{\log(3+2\sqrt2)}{2\sqrt2},
\label{groundstate}
\ear
which is, apart from a trivial shift of the whole spectrum due to different normalization, in agreement with the evaluation of the integral expression obtained  in \cite{MARTINS-SP2N} via the solution of the Bethe ansatz equations.

The remaining functions are given by,
\bear 
\omega_{II}^{(14)}(\lambda)&=&\frac{d}{d\lambda}\log\left[\frac{ \Gamma(\frac{7}{6}+\frac{\lambda}{3})\Gamma(\frac{5}{6}-\frac{\lambda}{3})}{\Gamma(\frac{7}{6}-\frac{\lambda}{3}) \Gamma(\frac{5}{6}+\frac{\lambda}{3})}\right] \nonumber \\
&+& \frac{d}{d\lambda}\log\left[\frac{ \Gamma(\frac{7}{16}-\frac{\lambda}{8})\Gamma(\frac{9}{16}+\frac{\lambda}{8}) \Gamma(\frac{15}{16}+\frac{\lambda}{8}) \Gamma(\frac{17}{16} -\frac{\lambda}{8})}
{  \Gamma(\frac{7}{16}+\frac{\lambda}{8}) \Gamma(\frac{9}{16}-\frac{\lambda}{8}) \Gamma(\frac{15}{16}-\frac{\lambda}{8}) \Gamma(\frac{17}{16} +\frac{\lambda}{8}) }\right] \nonumber\\
&+& \frac{d}{d\lambda}\log\left[\frac{ \Gamma(\frac{19}{16}+\frac{\lambda}{8}) \Gamma(\frac{23}{16}+\frac{\lambda}{8}) \Gamma(\frac{11}{16} - \frac{\lambda}{8}) \Gamma(\frac{15}{16}-\frac{\lambda}{8})}{ \Gamma(\frac{3}{16}-\frac{\lambda}{8}) \Gamma(\frac{7}{16}-\frac{\lambda}{8}) \Gamma(\frac{11}{16} + \frac{\lambda}{8}) \Gamma(\frac{15}{16}+\frac{\lambda}{8})} \right], \label{solw14} \\
\omega_{I}^{(6)}(\lambda)&=&-\omega_{II}^{(6)}(-4-\lambda), \\
\omega_{I}^{(14)}(\lambda)&=&-\omega_{II}^{(14)}(-4-\lambda),  \\
\omega^{(14')}(\lambda)&=&\frac{d}{d\lambda}\log{\left[\frac{\Gamma(\frac{11}{8}+\frac{\lambda}{8})\Gamma(\frac{7}{8}-\frac{\lambda}{8})}{\Gamma(\frac{3}{8}-\frac{\lambda}{8})\Gamma(\frac{7}{8}+\frac{\lambda}{8})} \right]}. \label{solw14p}
\ear

\begin{figure}[th]
	\begin{center}
		\includegraphics[width=0.65\linewidth, angle=-90]{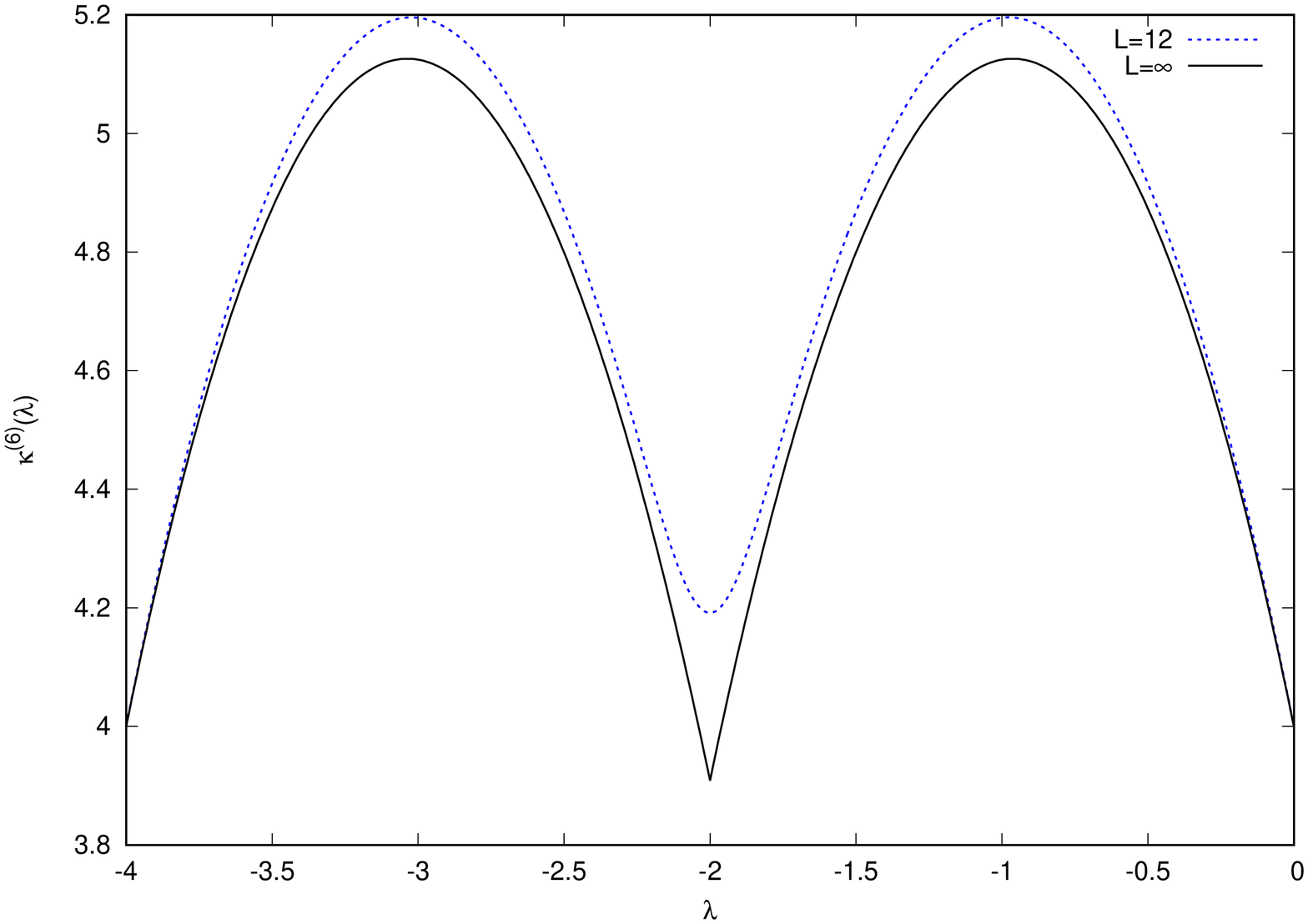}
		\caption{Partition function per site $\kappa^{(6)}_L(\lambda)=\left(\Lambda^{(6)}_0(\lambda)\right)^{1/L}$ as a function of $\lambda$ for finite horizontal lattice size $L=12$ and its comparison with the result in the thermodynamic limit ($L\rightarrow\infty$). Notice that $\kappa^{(6)}_L(\lambda)$ is analytic for $L$ finite but develops a cusp and is not analytic at $\lambda=-2$ for $L\to\infty$.}
		\label{fig2}
	\end{center}
\end{figure}

By integrating (\ref{solsp6},\ref{solw14},\ref{solw14p}) and fixing the integration constants such that the unitarity property is satisfied, we obtain the partition function per site given by,
\bear
\kappa_{II}^{(6)}(\lambda)&=&8^2  
\left[\frac{\Gamma(\frac{9}{8}+\frac{\lambda}{8})\Gamma(\frac{5}{8}-\frac{\lambda}{8})\Gamma(\frac{3}{2}+\frac{\lambda}{8})\Gamma(1-\frac{\lambda}{8})}{\Gamma(\frac{1}{8}-\frac{\lambda}{8})\Gamma(\frac{5}{8}+\frac{\lambda}{8})\Gamma(\frac{1}{2}-\frac{\lambda}{8})\Gamma(1+\frac{\lambda}{8})}\right] \left[\frac{\Gamma(\frac{1}{3}+\frac{\lambda}{3})\Gamma(\frac{2}{3}-\frac{\lambda}{3})}{ \Gamma(\frac{1}{3}-\frac{\lambda}{3})\Gamma(\frac{2}{3}+\frac{\lambda}{3})}\right] \nonumber \\
&\times&  \left[\frac{\Gamma(\frac{1}{8}-\frac{\lambda}{8})  \Gamma(\frac{1}{4}+\frac{\lambda}{8}) \Gamma(\frac{3}{8}-\frac{\lambda}{8})\Gamma(\frac{5}{8}+\frac{\lambda}{8})\Gamma(\frac{3}{4}-\frac{\lambda}{8})\Gamma(\frac{7}{8}+\frac{\lambda}{8})}{\Gamma(\frac{1}{8}+\frac{\lambda}{8})\Gamma(\frac{1}{4}-\frac{\lambda}{8})\Gamma(\frac{3}{8}+\frac{\lambda}{8})\Gamma(\frac{5}{8}-\frac{\lambda}{8})\Gamma(\frac{3}{4}+\frac{\lambda}{8})\Gamma(\frac{7}{8}-\frac{\lambda}{8})}\right], 
\ear
\bear
\kappa_{II}^{(14)}(\lambda)&=&8^2\left[\frac{ \Gamma(\frac{19}{16}+\frac{\lambda}{8}) \Gamma(\frac{23}{16}+\frac{\lambda}{8}) \Gamma(\frac{11}{16} - \frac{\lambda}{8}) \Gamma(\frac{15}{16}-\frac{\lambda}{8})}{ \Gamma(\frac{3}{16}-\frac{\lambda}{8}) \Gamma(\frac{7}{16}-\frac{\lambda}{8}) \Gamma(\frac{11}{16} + \frac{\lambda}{8}) \Gamma(\frac{15}{16}+\frac{\lambda}{8})} \right] \left[\frac{ \Gamma(\frac{7}{6}+\frac{\lambda}{3})\Gamma(\frac{5}{6}-\frac{\lambda}{3})}{\Gamma(\frac{7}{6}-\frac{\lambda}{3}) \Gamma(\frac{5}{6}+\frac{\lambda}{3})}\right] \nonumber \\
&\times&\left[\frac{ \Gamma(\frac{7}{16}-\frac{\lambda}{8})\Gamma(\frac{9}{16}+\frac{\lambda}{8}) \Gamma(\frac{15}{16}+\frac{\lambda}{8}) \Gamma(\frac{17}{16} -\frac{\lambda}{8})}
{  \Gamma(\frac{7}{16}+\frac{\lambda}{8}) \Gamma(\frac{9}{16}-\frac{\lambda}{8}) \Gamma(\frac{15}{16}-\frac{\lambda}{8}) \Gamma(\frac{17}{16} +\frac{\lambda}{8}) }\right], \label{solkappa14} \\
\kappa_{I}^{(6)}(\lambda)&=&\kappa_{II}^{(6)}(-4-\lambda),\quad \\ 
\kappa_{I}^{(14)}(\lambda)&=&\kappa_{II}^{(14)}(-4-\lambda),\quad \\
\kappa^{(14')}(\lambda)&=&8\left[\frac{\Gamma(\frac{11}{8}+\frac{\lambda}{8})\Gamma(\frac{7}{8}-\frac{\lambda}{8})}{\Gamma(\frac{3}{8}-\frac{\lambda}{8})\Gamma(\frac{7}{8}+\frac{\lambda}{8})} \right]. \label{solkappa14p}
\ear

\begin{figure}[tbh!]
	\begin{center}
		\includegraphics[width=0.65\linewidth, angle=-90]{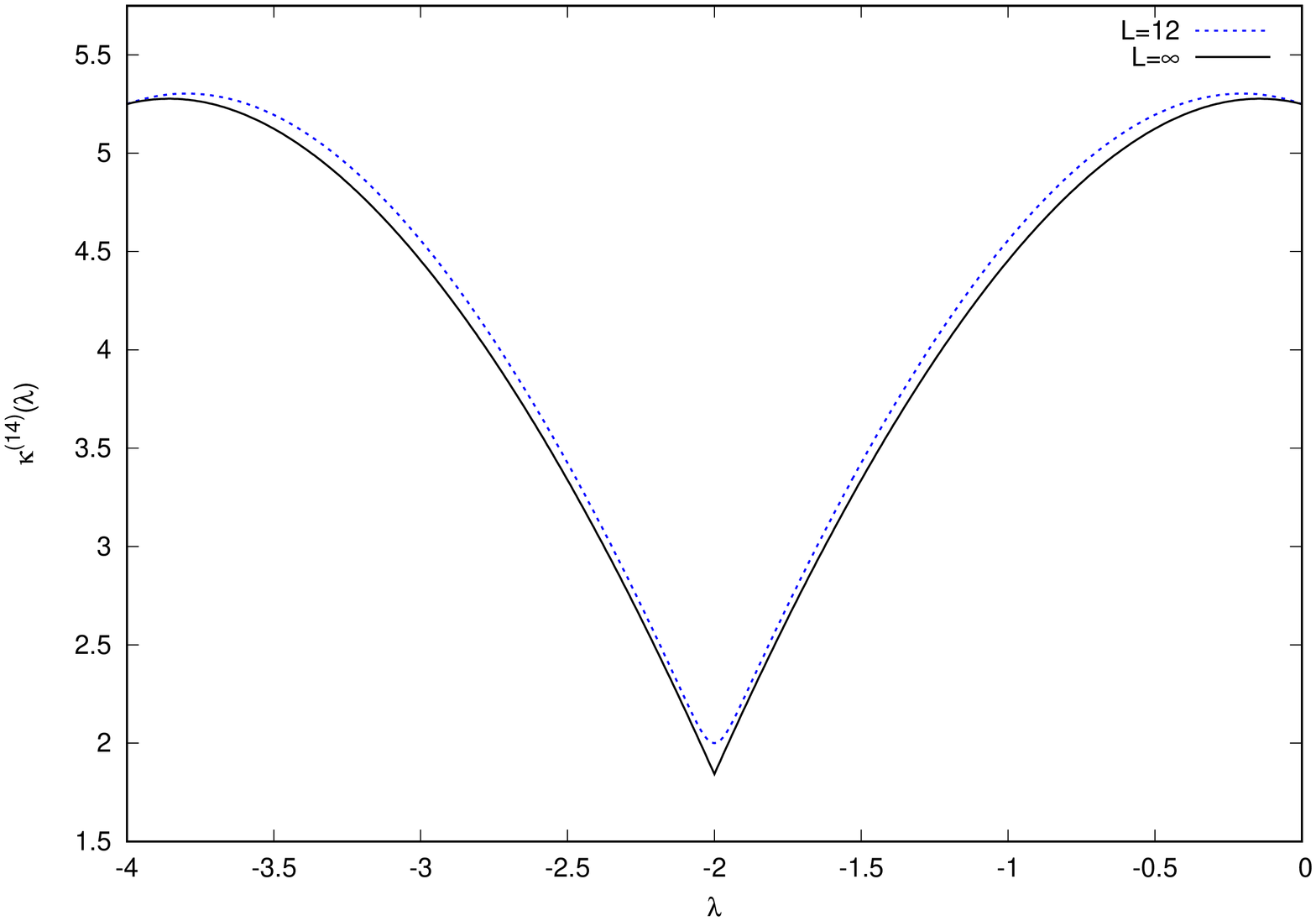}
		\caption{Partition function per site $\kappa^{(14)}_L(\lambda)=\left(\Lambda^{(14)}_0(\lambda)\right)^{1/L}$ as a function of $\lambda$ for finite horizontal lattice size $L=12$ and its comparison with the result in the thermodynamic limit ($L\rightarrow\infty$). Notice that $\kappa^{(14)}_L(\lambda)$ is analytic for $L$ finite but also develops a cusp and is not analytic at $\lambda=-2$ for $L\to\infty$.}
		\label{fig3}
	\end{center}
\end{figure}

\begin{figure}[th]
	\begin{center}
		\includegraphics[width=0.65\linewidth, angle=-90]{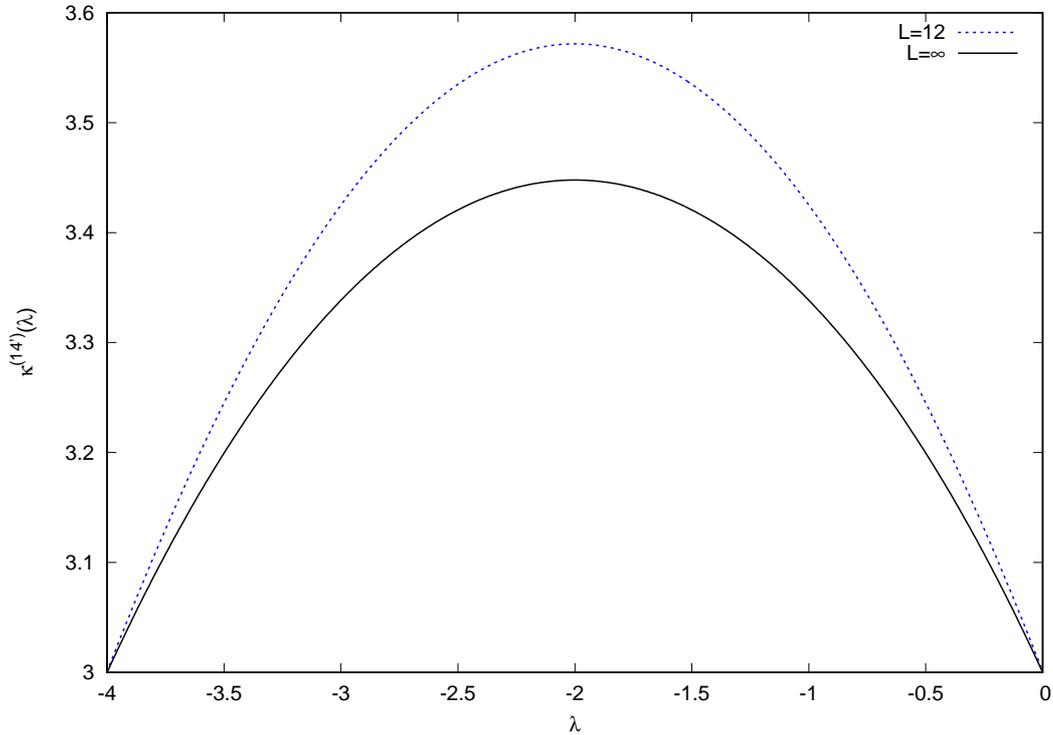}
		\caption{Partition function per site $\kappa^{(14')}_L(\lambda)=\left(\Lambda^{(14')}_0(\lambda)\right)^{1/L}$ as a function of $\lambda$ for finite horizontal lattice size $L=12$ and its comparison with the result in the thermodynamic limit ($L\rightarrow\infty$).}
		\label{fig4}
	\end{center}
\end{figure}

We show in Figures \ref{fig2}, \ref{fig3} and \ref{fig4} the comparison of the partition functions per site  $\kappa^{(\alpha)}_L(\lambda)=\left(\Lambda^{(\alpha)}_0(\lambda)\right)^{1/L}$ for $\alpha=6,14$ and $14'$ for the finite horizontal lattice size $L=12$ as a function of $\lambda$ with the result in the thermodynamic limit.

\newpage

\section{The general $Sp(2n)$ case}\label{general}

For the $Sp(2n)$ case, the tensor product of two fundamental representations decomposes as $2n\otimes 2n = 1 \oplus (2n+1)(n-1) \oplus (2n+1)n$ \cite{GROUP}. Therefore, the fundamental $R$-matrix $R_{12}^{(2n,2n)}(\lambda)$ defined in (\ref{Rmatrix}) can be rewritten as,
\eq
R^{(2n,2n)}_{12}(\lambda)=(\lambda+1)(\lambda-\Delta) \check{P}_{12}^{(1)} + (\lambda-1)(\lambda+\Delta) \check{P}_{12}^{((2n+1)(n-1))}  + (\lambda+1)(\lambda+\Delta) \check{P}_{12}^{((2n+1)n)},
\label{r2n2n}  
\en
where the projectors $\check{P}_{12}^{(\alpha)}$ are given in the Appendix~B. This again shows explicitly the singular values that degenerate in projection operators, which can be exploited to derive the fusion hierarchy recursively (see \cite{CAO} and the Appendix~B for more details on the fusion hierarchy for $Sp(2n)$ case).

Nevertheless, it is not convenient for the general $Sp(2n)$ case to label the spaces on which the $R$-matrices act in terms of the dimension of the representation. Instead, we choose to label it in terms of the Dynkin labels $[r_1,r_2,\dots,r_n]$ of the representations. Actually, we define a shorthand notation to indicate the fundamental representations such that the representation of dimension $2n$ is indicated by $\{1\}:=[1,0,\dots,0]$. We proceed similarly for the all other fundamental representations such that  $\{k\}:=[0,\dots,0,\overbrace{1}^{k-th},0,\dots,0]$, for $k=1,2,\dots,n$. For instance, the transfer matrix $T^{(\{1\})}(\lambda):=T^{(2n)}(\lambda)$ and so on (see Appendix~B for more details).

The transfer matrix fusion identities can be extended to the general $Sp(2n)$ along the same lines as done for the cases $Sp(4)$ \cite{RKP} and $Sp(6)$ in section \ref{sp6}. The final relations are given by,
\bear
T^{(\{1\})}(\lambda)T^{(\{1\})}(\lambda-\Delta)&=& [(\lambda^2-1) (\lambda^2-\Delta^2)]^L I  (1 + O(e^{-L})), \label{TMFISp2n1} \nonumber\\
T^{(\{1\})}(\lambda)T^{(\{m\})}(\lambda-\tfrac{m+1}{2})&=& [(\lambda-1) (\lambda+\Delta)]^L T^{(\{m+1\})}(\lambda-\tfrac{m}{2}) (1 + O(e^{-L})), \label{TMFISp2n2} \nonumber\\
T^{(\{1\})}(\lambda)T^{(\{n-1\})}(\lambda-\tfrac{n}{2})&=& [(\lambda^2-1) (\lambda+\Delta)]^L T^{(\{n\})}(\lambda-\tfrac{n-1}{2})(1 + O(e^{-L})), \label{TMFISp2n3}\\
T^{(\{1\})}(\lambda)T^{(\{m+1\})}(\lambda-\tfrac{2\Delta-m}{2})&=& [(\lambda-1)(\lambda+\Delta)]^L T^{(\{m\})}(\lambda-\tfrac{2\Delta-m-1}{2}) (1 + O(e^{-L})), \nonumber\\
T^{(\{1\})}(\lambda)T^{(\{n\})}(\lambda-\tfrac{\Delta+2}{2})&=& [(\lambda+\Delta)]^L T^{(\{n-1\})}(\lambda-\tfrac{\Delta+1}{2}) (1 + O(e^{-L})), \label{TMFISp2n5}\nonumber
\ear
for $m=1,2,\dots,n-2$. 

Similar relations hold for all the fused transfer matrix eigenvalues, which include the leading eigenvalues denoted $\Lambda_0^{(\{m\})}(\lambda)$ for $m=1,\dots,n$.  The partition function per site in the thermodynamic limit is defined as $\kappa^{(m)}(\lambda)= \lim_{L\rightarrow\infty} \left(\Lambda_0^{(\{m\})}(\lambda)\right)^{1/L}$. Consequently, the above relations can be rewritten in terms of the partition function $\kappa^{(m)}(\lambda)$ as follows,
\bear
\kappa_{II}^{(1)}(\lambda)\kappa_{I}^{(1)}(\lambda-\Delta)&=& (\lambda^2-1) (\lambda^2-\Delta^2), \nonumber\\
\kappa_{II}^{(1)}(\lambda)\kappa_{II}^{(m)}(\lambda-\tfrac{m+1}{2})&=& (\lambda-1) (\lambda+\Delta) \kappa_{II}^{(m+1)}(\lambda-\tfrac{m}{2}),  \nonumber \\
\kappa_{II}^{(1)}(\lambda)\kappa_{II}^{(n-1)}(\lambda-\tfrac{n}{2})&=& (\lambda^2-1) (\lambda+\Delta) \kappa^{(n)}(\lambda-\tfrac{n-1}{2}), \label{kappaeqssp2n}\\
\kappa_{II}^{(1)}(\lambda)\kappa_{I}^{(m+1)}(\lambda-\tfrac{2\Delta-m}{2})&=& (\lambda-1)(\lambda+\Delta) \kappa_{I}^{(m)}(\lambda-\tfrac{2\Delta-m-1}{2}), \nonumber \\
\kappa_{II}^{(1)}(\lambda)\kappa^{(n)}(\lambda-\tfrac{\Delta+2}{2})&=& (\lambda+\Delta) \kappa_{I}^{(n-1)}(\lambda-\tfrac{\Delta+1}{2}), \nonumber
\ear
for $m=1,2,\dots,n-2$. Based on the studies for the $Sp(4)$ \cite{RKP} and $Sp(6)$ cases, we infer at this point that only the leading eigenvalue  of last fusion level $\Lambda^{ (\{n\})}_0(\lambda)$ is free of zeros inside the analyticity strip, which is assumed to be $-\frac{1}{2}-\Delta<\mbox{Re}(\lambda)<\frac{1}{2}$. We also assume that all other eigenvalues have zeros at the center of the strip distributed along an infinitely long vertical line. Therefore, the indices $I$ and $II$ specify the functions on the left and right of the cut line $\mbox{Re}(\lambda)=-\frac{\Delta}{2}$.

By taking the logarithmic derivative of the partition function, we introduce the functions $\omega^{(m)}(\lambda)=\tfrac{d}{d\lambda}\log\kappa^{(m)}(\lambda)$ for $m=1,\dots,n$. This allows us to rewrite the Eqs.(\ref{kappaeqssp2n}) as given below,
\bear
\omega_{II}^{(1)}(\lambda)+\omega_{I}^{(1)}(\lambda-\Delta)&=& \frac{1}{\lambda-1} + \frac{1}{\lambda+1} + \frac{1}{\lambda-\Delta} + \frac{1}{\lambda+\Delta}, \nonumber\\
\omega_{II}^{(1)}(\lambda)+\omega_{II}^{(m)}(\lambda-\tfrac{m+1}{2})&=& \frac{1}{\lambda-1} + \frac{1}{\lambda+\Delta} + \omega_{II}^{(m+1)}(\lambda-\tfrac{m}{2}),   \nonumber \\
\omega_{II}^{(1)}(\lambda)+\omega_{II}^{(n-1)}(\lambda-\tfrac{n}{2})&=& \frac{1}{\lambda-1} + \frac{1}{\lambda+1} +\frac{1}{\lambda+\Delta} +\omega^{(n)}(\lambda-\tfrac{n-1}{2}), \label{eqsomegaSp2n} \\
\omega_{II}^{(1)}(\lambda)+\omega_{I}^{(m+1)}(\lambda-\tfrac{2\Delta-m}{2})&=& \frac{1}{\lambda-1}+ \frac{1}{\lambda+\Delta} + \omega_{I}^{(m)}(\lambda-\tfrac{2\Delta-m-1}{2}), \nonumber \\
\omega_{II}^{(1)}(\lambda)+\omega^{(n)}(\lambda-\tfrac{\Delta+2}{2})&=& \frac{1}{\lambda+\Delta} + \omega_{I}^{(n-1)}(\lambda-\tfrac{\Delta+1}{2}), \nonumber
\ear
for $m=1,2,\dots,n-2$.

Similar to the cases $n=2$ and $3$, a simpler equation can be obtained from the Eqs.(\ref{eqsomegaSp2n}) by elimination,
\bear
\omega^{(n)}(\lambda-\tfrac{n-1}{2})+\omega^{(n)}(\lambda-\tfrac{3\Delta-2}{2}) = \frac{1}{\lambda+2} + \frac{1}{\lambda-\Delta},
\ear
which is in agreement with (\ref{omega14pFuncEq}) for $n=3$.

The solution for the arbitrary $Sp(2n)$ is therefore obtained by Fourier-Laplace transform and the final result for the fundamental representation ($m=1$) is conveniently written in terms of gamma functions and an integral expression as follows,
\bear
&&\omega_{II}^{(1)}(\lambda)=\int_{0}^{\infty}\frac{e^{-n t}(1-e^{-t})}{(1-e^{-n t})(1+e^{-(n+1)t})}\left( e^{ \lambda t} + e^{- \lambda t}\right) dt  \label{omegaSp2n1} \\
&+& \frac{d}{d\lambda}\log\left[\frac{\Gamma(\frac{2n+3}{2(n+1)}+\frac{\lambda}{2(n+1)}) \Gamma(\frac{n+2}{2(n+1)}-\frac{\lambda}{2(n+1)}) \Gamma(\frac{3}{2}+\frac{\lambda}{2(n+1)})  \Gamma(1-\frac{\lambda}{2(n+1)})}{\Gamma(\frac{1}{2(n+1)}-\frac{\lambda}{2(n+1)}) \Gamma(\frac{n+2}{2(n+1)}+\frac{\lambda}{2(n+1)}) \Gamma(\frac{1}{2}-\frac{\lambda}{2(n+1)})  \Gamma(1+\frac{\lambda}{2(n+1)})} \right]. \nonumber
\ear
The homogeneous limit ($\lambda=0$) of the above function results precisely in the ground state energy of the quantum spin chain, which is, apart from a trivial shift of the whole spectrum by $\tfrac{1}{\Delta}$ due to different normalization, in agreement with the integral expression obtained via the solution of the Bethe ansatz equations in \cite{MARTINS-SP2N} as verified by the numerical evaluation of the integrals for $n>3$. It is worth noting that the last term of (\ref{omegaSp2n1}), which is written in terms of the gamma functions, would be the solution of the first equation in (\ref{eqsomegaSp2n}) if the eigenvalue expression was free of zeros inside the strip. 
Consequently, the first term in (\ref{omegaSp2n1}) can be seen as a kind of CDD factor due to the break in the analyticity properties at the center of the strip. In the general $Sp(2n)$ case, we could not rewrite this integral in terms of gamma function, since the partial fraction expansion of the integrand needed in this process changes greatly with the values of $n$.

The general solution for the remaining functions is written as follows,
\bear
&&\omega_{II}^{(m)}(\lambda)=\int_{0}^{\infty}\frac{e^{-(\frac{2n-m+1}{2}) t}(1-e^{-m t})}{(1-e^{-n t})(1+e^{-(n+1)t})}\left( e^{ \lambda t} + e^{- \lambda t}\right) dt  \label{omegaSp2nm} \\
&+& \frac{d}{d\lambda}\log\left[\frac{\Gamma(\frac{4n+m+5}{4(n+1)}+\frac{\lambda}{2(n+1)})  \Gamma(\frac{2n+m+3}{4(n+1)}-\frac{\lambda}{2(n+1)}) \Gamma(\frac{6n-m+7}{4(n+1)}+\frac{\lambda}{2(n+1)})  \Gamma(\frac{4n-m+5}{4(n+1)}-\frac{\lambda}{2(n+1)})}{\Gamma(\frac{m+1}{4(n+1)}-\frac{\lambda}{2(n+1)})  \Gamma(\frac{2n+m+3}{4(n+1)}+\frac{\lambda}{2(n+1)}) \Gamma(\frac{2n-m+3}{4(n+1)}-\frac{\lambda}{2(n+1)})  \Gamma(\frac{4n-m+5}{4(n+1)}+\frac{\lambda}{2(n+1)})} \right], \nonumber
\ear
for $m=2,\dots,n-1$ and
\bear
\omega^{(n)}(\lambda)&=&
\frac{d}{d\lambda}\log\left[\frac{\Gamma(\frac{5n+7}{4(n+1)}+\frac{\lambda}{2(n+1)}) \Gamma(\frac{3n+5}{4(n+1)}-\frac{\lambda}{2(n+1)}) }{\Gamma(\frac{n+3}{4(n+1)}-\frac{\lambda}{2(n+1)}) \Gamma(\frac{3n+5}{4(n+1)}+\frac{\lambda}{2(n+1)}) } \right]. \label{omegaSp2nn}
\ear

Finally, by integrating (\ref{omegaSp2n1}--\ref{omegaSp2nn}) and fixing the integration constants such that the unitarity property is satisfied, we finally obtain the partition function per site given by,
\begin{alignat}{1}
&\kappa_{II}^{(1)}(\lambda)=(2(n+1))^2 \exp\left\{\int_{0}^{\infty}\frac{e^{-n t}(1-e^{-t})}{t(1-e^{-n t})(1+e^{-(n+1)t})}\left( e^{ \lambda t} - e^{- \lambda t}\right) dt \right\}  \label{logkappaSp2n1} \\
&\times \left[\frac{\Gamma(\frac{2n+3}{2(n+1)}+\frac{\lambda}{2(n+1)}) \Gamma(\frac{n+2}{2(n+1)}-\frac{\lambda}{2(n+1)}) \Gamma(\frac{3}{2}+\frac{\lambda}{2(n+1)})  \Gamma(1-\frac{\lambda}{2(n+1)})}{\Gamma(\frac{1}{2(n+1)}-\frac{\lambda}{2(n+1)}) \Gamma(\frac{n+2}{2(n+1)}+\frac{\lambda}{2(n+1)}) \Gamma(\frac{1}{2}-\frac{\lambda}{2(n+1)})  \Gamma(1+\frac{\lambda}{2(n+1)})} \right], \nonumber \\
&\kappa_{II}^{(m)}(\lambda)=(2(n+1))^2 \exp\left\{\int_{0}^{\infty}\frac{e^{-(\frac{2n-m+1}{2}) t}(1-e^{-m t})}{t(1-e^{-n t})(1+e^{-(n+1)t})}\left( e^{ \lambda t} - e^{- \lambda t}\right) dt \right\}   \\
&\times \left[\frac{\Gamma(\frac{4n+m+5}{4(n+1)}+\frac{\lambda}{2(n+1)})  \Gamma(\frac{2n+m+3}{4(n+1)}-\frac{\lambda}{2(n+1)}) \Gamma(\frac{6n-m+7}{4(n+1)}+\frac{\lambda}{2(n+1)})  \Gamma(\frac{4n-m+5}{4(n+1)}-\frac{\lambda}{2(n+1)})}{\Gamma(\frac{m+1}{4(n+1)}-\frac{\lambda}{2(n+1)})  \Gamma(\frac{2n+m+3}{4(n+1)}+\frac{\lambda}{2(n+1)}) \Gamma(\frac{2n-m+3}{4(n+1)}-\frac{\lambda}{2(n+1)}) \Gamma(\frac{4n-m+5}{4(n+1)}+\frac{\lambda}{2(n+1)})} \right], \nonumber
\end{alignat}
for $m=2,\dots,n-1$ and
\begin{alignat}{1}
&\kappa^{(n)}(\lambda)=
2(n+1)\left[\frac{\Gamma(\frac{5n+7}{4(n+1)}+\frac{\lambda}{2(n+1)}) \Gamma(\frac{3n+5}{4(n+1)}-\frac{\lambda}{2(n+1)}) }{\Gamma(\frac{n+3}{4(n+1)}-\frac{\lambda}{2(n+1)}) \Gamma(\frac{3n+5}{4(n+1)}+\frac{\lambda}{2(n+1)}) } \right]. 
\end{alignat}

\newpage

\section{Conclusion}\label{CONCLUSION}

We investigated the partition function of the fundamental $Sp(2n)$ vertex model on a square lattice in the thermodynamic limit via a functional approach. 

This was done in great detail for the special case of the $Sp(6)$ vertex model, which together with the previous results for the $Sp(4)$ case \cite{RKP} allowed for the generalization for the arbitrary $Sp(2n)$ case. This is a subtle calculation, since all fusion level leading eigenvalues but the last have a vertical line of zeros at the center of the analyticity strip. In the thermodynamic limit, this vertical line of zeros becomes an extended singularity, which divides the complex plane into two parts. The established transfer matrix fusion relations are used to connect both sides of the analyticity strip, which therefore allows for the computation of the partition function per site of the fundamental representation of the $Sp(2n)$ vertex model. In addition, we also obtained the partition function of vertex models mixing different representations. 

We expect that the determination of the partition function is an important step toward the calculation of the two-sites correlation functions of the $Sp(2n)$ quantum spin chain, along the same lines as done for the $O(n)$ spin chain \cite{RIBEIRO}. However, there still exists the challenge of the derivation of fused quantum Knizhnik-Zamolodchikov equations, which should completely determine the correlations. We also expect that the method used in this work can be extended to other models with similar analytic subtleties.

\section*{Acknowledgments}

GAPR thanks A. Kl\"umper and P.A. Pearce for the collaboration in the previous stage of this project and M.J. Martins for discussions.

\section*{Appendix A: $Sp(6)$ fusion rules}

In this appendix, we present more details concerning fusion in the $Sp(6)$ case.

The projectors $\check{P}_{12}^{(\alpha)}$ for $\alpha=1,14,21$ which arise from the decomposition $6\otimes 6= 1\oplus 14 \oplus 21$ \cite{GROUP} are simply related to the identity, permutation and Temperley-Lieb operators already defined. Therefore, we just list their explicit relations as follows,
\begin{align}
\check{P}_{12}^{(1)}=-\frac{1}{6} E_{12}, \quad \check{P}_{12}^{(14)}= \frac{1}{2} \left( I_{12} -  P_{12} \right) + \frac{1}{6} E_{12}, \quad 	\check{P}_{12}^{(21)}= \frac{1}{2}\left( I_{12} +  P_{12} \right). 
\end{align}

On the other hand, the projectors $\check{P}_{12}^{(6)}$, $\check{P}_{12}^{(14')}$ and $\check{P}_{12}^{(64)}$  are due to the decomposition $14\otimes 6 = 6 \oplus 14' \oplus 64 $. As the projectors are constrained by the usual relation  $\check{P}_{12}^{(6)}+\check{P}_{12}^{(14')}+\check{P}_{12}^{(64)}=I$, we only list the projectors on the $6$ and $14'$-dimensional spaces,
\bear
\check{P}_{12}^{(6)}=\sum_{i=1}^6 \ket{\phi^{(6)}_i}\bra{\phi^{(6)}_i},
\ear
\bear
\ket{\phi^{(6)}_1}&=& \tfrac{\sqrt3}{14}( \ket{1,5} + \ket{2,4}  - \ket{4,3} - \ket{6,2} )  +\tfrac{1}{2} \tfrac{\sqrt{3}}{7} \ket{7,1} -\tfrac{1}{2\sqrt7} \ket{10,1}, \nonumber \\
\ket{\phi^{(6)}_2}&=& \tfrac{\sqrt3}{14}( -\ket{1,6} + \ket{3,4} - \ket{5,3} - \ket{11,1} ) - \tfrac{1}{2} \tfrac{\sqrt3}{7} \ket{7,2} -\tfrac{1}{2\sqrt7} \ket{10,2}, \nonumber \\   
\ket{\phi^{(6)}_3}&=& -\tfrac{\sqrt3}{14}(  \ket{2,6} + \ket{3,5} + \ket{8,2} + \ket{12,1}) +  \tfrac{1}{\sqrt7} \ket{10,3}, \nonumber \\
\ket{\phi^{(6)}_4}&=& -\tfrac{\sqrt3}{14}(  \ket{4,6} + \ket{5,5} + \ket{9,2} + \ket{13,1} ) +  \tfrac{1}{\sqrt7} \ket{10,4} ),  \\
\ket{\phi^{(6)}_5}&=& \tfrac{\sqrt3}{14}( -\ket{6,6} - \ket{8,4} + \ket{9,3} - \ket{14,1} ) - \tfrac{1}{2} \tfrac{\sqrt3}{7} \ket{7,5}  -\tfrac{1}{2\sqrt7} \ket{10,5}, \nonumber \\
\ket{\phi^{(6)}_6}&=& \tfrac{\sqrt3}{14}( \ket{11,5} + \ket{12,4} - \ket{13,3} - \ket{14,2} ) -\tfrac{1}{2} \tfrac{\sqrt3}{7} \ket{7,6} + \tfrac{1}{2\sqrt7} \ket{10,6}, \nonumber
\ear

\bear
\check{P}_{12}^{(14')}=\sum_{i=1}^{14} \ket{\phi^{(14')}_i}\bra{\phi^{(14')}_i},
\ear
\bear
\ket{\phi^{(14')}_1}&=& \tfrac{1}{\sqrt3}( -\ket{1,3} + \ket{2,2}  - \ket{3,1}), \nonumber \\ 
\ket{\phi^{(14')}_2}&=& \tfrac{1}{\sqrt3}( -\ket{1,4} + \ket{4,2}  - \ket{5,1}), \nonumber \\ 
\ket{\phi^{(14')}_3}&=& \tfrac{1}{\sqrt3}( -\ket{2,5} + \ket{6,3}  - \ket{8,1}), \nonumber \\ 
\ket{\phi^{(14')}_4}&=& \tfrac{1}{\sqrt3}( -\ket{4,5} + \ket{6,4}  - \ket{9,1}), \nonumber \\ 
\ket{\phi^{(14')}_5}&=& \tfrac{1}{\sqrt6}( -\ket{1,5} + \ket{2,4} - \ket{4,3} +\ket{6,2})  - \tfrac{1}{2\sqrt3} \ket{7,1}  -\tfrac{1}{2}\ket{10,1}, \nonumber \\
\ket{\phi^{(14')}_6}&=& \tfrac{1}{\sqrt6}( - \ket{1,6}  - \ket{3,4} + \ket{5,3} - \ket{11,1} )  - \tfrac{1}{2\sqrt3} \ket{7,2}  + \tfrac{1}{2} \ket{10,2}, \nonumber \\  
\ket{\phi^{(14')}_7}&=& \tfrac{1}{\sqrt6}( - \ket{2,6} + \ket{3,5} + \ket{8,2} - \ket{12,1} )  - \tfrac{1}{\sqrt3} \ket{7,3}, \nonumber \\
\ket{\phi^{(14')}_8}&=& \tfrac{1}{\sqrt3}(  - \ket{3,6} + \ket{11,3} - \ket{12,2} ),  \\ 
\ket{\phi^{(14')}_9}&=& \tfrac{1}{\sqrt6}( - \ket{4,6} + \ket{5,5} + \ket{9,2} - \ket{13,1} )  - \tfrac{1}{\sqrt3} \ket{7,4}, \nonumber \\
\ket{\phi^{(14')}_{10}}&=& \tfrac{1}{\sqrt3}(  - \ket{5,6}  + \ket{11,4} - \ket{13,2} ),\nonumber \\ 
\ket{\phi^{(14')}_{11}}&=& \tfrac{1}{\sqrt6}( - \ket{6,6}  + \ket{8,4} - \ket{9,3} - \ket{14,1} )  - \tfrac{1}{2\sqrt3} \ket{7,5}  + \tfrac{1}{2} \ket{10,5}, \nonumber \\ 
\ket{\phi^{(14')}_{12}}&=& \tfrac{1}{\sqrt6}(- \ket{12,4} + \ket{11,5}  + \ket{13,3} - \ket{14,2} )  - \tfrac{1}{2\sqrt3} \ket{7,6}  -\tfrac{1}{2} \ket{10,6}, \nonumber \\ 
\ket{\phi^{(14')}_{13}}&=& \tfrac{1}{\sqrt3}(- \ket{8,6}  + \ket{12,5} - \ket{14,3} ), \nonumber \\
\ket{\phi^{(14')}_{14}}&=& \tfrac{1}{\sqrt3}(- \ket{9,6}  + \ket{13,5} - \ket{14,4}). \nonumber 
 \ear

Finally, the tensor product decomposition of $14'\otimes 6= 14 \oplus 70$ introduces another $(14\times 6)$-dimensional projector, but now on the spaces $14$ and $70$-dimensional given by $\check{P'}_{12}^{(14)}$ and $\check{P}_{12}^{(70)}$. Again, due to the relation  $\check{P'}_{12}^{(14)}+\check{P}_{12}^{(70)}=I$, we list only the projector on the $14$-dimensional space,
\bear
\check{P'}_{12}^{(14)}=\sum_{i=1}^{14} \ket{\phi^{(14)}_i}\bra{\phi^{(14)}_i},
\ear
\bear
\ket{\phi^{(14)}_1}&=& \tfrac{1}{\sqrt3}( \ket{1,4} - \ket{2,3} )  -\tfrac{1}{\sqrt6}( \ket{5,2} + \ket{6,1} ), \nonumber \\ 
\ket{\phi^{(14)}_2}&=& -\tfrac{1}{\sqrt3}( \ket{1,5}+ \ket{3,2} )  + \tfrac{1}{\sqrt6} (\ket{5,3} - \ket{7,1} ), \nonumber \\
\ket{\phi^{(14)}_3}&=& \tfrac{1}{\sqrt3}( \ket{1,6}  - \ket{8,1} )  + \tfrac{1}{\sqrt6} ( \ket{7,2} - \ket{6,3} ), \nonumber \\ 
\ket{\phi^{(14)}_4}&=& -\tfrac{1}{\sqrt3}( \ket{2,5} + \ket{4,2} )  + \tfrac{1}{\sqrt6} ( \ket{5,4} - \ket{9,1} ), \nonumber \\
\ket{\phi^{(14)}_5}&=& \tfrac{1}{\sqrt3}( \ket{2,6}  - \ket{10,1} )  + \tfrac{1}{\sqrt6} ( \ket{9,2} - \ket{6,4} ), \nonumber \\ 
\ket{\phi^{(14)}_6}&=& -\tfrac{1}{\sqrt3}( \ket{3,4} - \ket{4,3} )  - \tfrac{1}{\sqrt6} ( \ket{5,5} + \ket{11,1}  ), \nonumber \\ 
\ket{\phi^{(14)}_7}&=& \tfrac{1}{2}( \ket{6,5} + \ket{7,4} - \ket{9,3} - \ket{11,2} ), \\
\ket{\phi^{(14)}_8}&=& \tfrac{1}{2}( \ket{5,6} + \ket{7,4}  -  \ket{9,3} - \ket{12,1}), \nonumber \\
\ket{\phi^{(14)}_9}&=& \tfrac{1}{\sqrt3}(  \ket{8,4}  - \ket{10,3} )  - \tfrac{1}{\sqrt6} ( \ket{12,2} + \ket{6,6} ), \nonumber \\
\ket{\phi^{(14)}_{10}}&=& \tfrac{1}{\sqrt3}(  \ket{3,6}  - \ket{13,1} )  + \tfrac{1}{\sqrt6} ( \ket{11,3} - \ket{7,5} ), \nonumber \\
\ket{\phi^{(14)}_{11}}&=& -\tfrac{1}{\sqrt3}( \ket{8,5}  + \ket{13,2})  + \tfrac{1}{\sqrt6} ( \ket{12,3} - \ket{7,6} ), \nonumber \\  
\ket{\phi^{(14)}_{12}}&=& \tfrac{1}{\sqrt3}(  \ket{4,6}  - \ket{14,1} )  + \tfrac{1}{\sqrt6} ( \ket{11,4} - \ket{9,5} ), \nonumber \\
\ket{\phi^{(14)}_{13}}&=& -\tfrac{1}{\sqrt3}( \ket{10,5} + \ket{14,2} )  + \tfrac{1}{\sqrt6} ( \ket{12,4} - \ket{9,6} ), \nonumber \\
\ket{\phi^{(14)}_{14}}&=& \tfrac{1}{\sqrt3}(  \ket{13,4}  - \ket{14,3} )  + \tfrac{1}{\sqrt6} ( \ket{12,5} + \ket{11,6} ). \nonumber 
\ear

By exploiting the singular values of $R_{12}^{(6,6)}(\lambda)$ we have that,
\bear
\check{P}_{ab}^{(1)} R_{b2}^{(6,6)}(\lambda)R_{a2}^{(6,6)}(\lambda-4)\check{P}_{ab}^{(1)} &=&(\lambda^2-1)(\lambda^2-4^2)\check{P}_{ab}^{(1)},  \label{fusSp61} \\
\check{P}_{ab}^{(14)}R_{b2}^{(6,6)}(\lambda)R_{a2}^{(6,6)}(\lambda-1)\check{P}_{ab}^{(14)}&=&(\lambda-1)(\lambda+4) ~ R_{12}^{(14,6)}(\lambda-\frac{1}{2}). \label{fusSp62}
\ear

Again, we can use the singular points of $R_{12}^{(14,6)}(\lambda)$ to obtain,
\bear
\check{P}_{ab}^{(14')} R_{b2}^{(6,6)}(\lambda)R_{a2}^{(14,6)}(\lambda-\frac{3}{2})\check{P}_{ab}^{(14')}&=&(\lambda^2-1)(\lambda+4) R_{12}^{(14',6)}(\lambda-1), \label{fusSp63}  \\
\check{P}_{ab}^{(6)} R_{b2}^{(6,6)}(\lambda)R_{a2}^{(14,6)}(\lambda-\frac{7}{2})\check{P}_{ab}^{(6)}&=&(\lambda-1)(\lambda+4) R_{12}^{(6,6)}(\lambda-3). \label{fusSp64}
\ear

Finally, we have one last fusion relation to close the set of fusion relations, namely
\bear
\check{P'}_{ab}^{(14)} R_{b2}^{(6,6)}(\lambda)R_{a2}^{(14',6)}(\lambda-3)\check{P'}_{ab}^{(14)}&=&(\lambda+4) R_{12}^{(14,6)}(\lambda-\frac{5}{2}). \label{fusSp65}  
\ear

The above relations can be naturally extended to the product of monodromy matrices, ${\cal T}_{\cal A}^{(\alpha,6)}(\lambda)=R_{{\cal A}L}^{(\alpha,6)}(\lambda)\dots R_{{\cal A}1}^{(\alpha,6)}(\lambda)$, with \mbox{$\alpha=6,14,14'$}. Therefore, the transfer matrix fusion relations (\ref{TMFRSp61}-\ref{TMFRSp64}) are naturally obtained from the fusion relations. 

For instance, by inserting the identity as the sum of the projectors into the trace, moving them around the trace and finally, by using (\ref{fusSp61}) we see that,
\bear
&&T^{(6)}(\lambda)T^{(6)}(\lambda-4)=\tr_{a \otimes b}{\left[ {\cal T}_{b}^{(6,6)}(\lambda) {\cal T}_{a}^{(6,6)}(\lambda-4)\right]}, \nonumber \\
&=& \tr_{a \otimes b}{\left[ \left(P_{ab}^{(1)} + P_{ab}^{(14)} +P_{ab}^{(21)}  \right) {\cal T}_{b}^{6,6)}(\lambda) {\cal T}_{a}^{(6,6)}(\lambda-4)\right]}, \\
&=& \tr_{a \otimes b}{\left[ P_{ab}^{(1)}  {\cal T}_{b}^{(6,6)}(\lambda) {\cal T}_{a}^{(6,6)}(\lambda-4) P_{ab}^{(1)} \right]} \nonumber\\
&+& \sum_{\alpha=14,21}\tr_{a \otimes b}{\left[ P_{ab}^{(\alpha)}  {\cal T}_{b}^{(6,6)}(\lambda) {\cal T}_{a}^{(6,6)}(\lambda-4) P_{ab}^{(\alpha)} \right]}, \nonumber \\
&=& \left[(\lambda^2-1)(\lambda^2-4^2)\right]^L I \nonumber\\
&+& \sum_{\alpha=14,21}\tr_{a \otimes b}{\left[ P_{ab}^{(\alpha)}  {\cal T}_{b}^{(6,6)}(\lambda) {\cal T}_{a}^{(6,6)}(\lambda-4) P_{ab}^{(\alpha)} \right]}, \nonumber
\ear
which gives the transfer matrix inversion identity (\ref{TMFISp61}), where the additional terms indicated by the sum  $\alpha=14,21$ encompass terms that are exponentially small in the thermodynamic limit.  It is worth noting that the above inversion relation at $\lambda=0$ is exact for arbitrary length $L$ due to the product of projection operators on different subspaces.  The remaining fusion relations are obtained along the same lines as above.

For latter convenience, we list the tensor product decomposition represented in terms of the dimensions of the irreducible representation and in terms of the Dynkin labels of the representation.
\begin{alignat}{3}
&6\otimes 6&= 1\oplus 14 \oplus 21, \quad \mbox{or} \quad [1,0,0]\otimes[1,0,0]&=[0,0,0]\oplus [0,1,0] \oplus[2,0,0], \nonumber \\ 
&6\otimes 14&= 6\oplus 14' \oplus 64, \quad \mbox{or} \quad [1,0,0]\otimes[0,1,0]&=[1,0,0]\oplus [0,0,1] \oplus[1,1,0], \\
&6\otimes 14'&= 14 \oplus 70, \quad \mbox{or} \quad [1,0,0]\otimes[0,0,1]&=[0,1,0]\oplus [1,0,1]. \nonumber
\end{alignat}

\section*{Appendix B: $Sp(2n)$ fusion rules}

The projectors $\check{P}_{12}^{(\alpha)}$ for $\alpha=1,(2n+1)(n-1),(2n+1)n$ which arise from the decomposition $2n\otimes 2n= 1\oplus (2n+1)(n-1) \oplus (2n+1)n$ \cite{GROUP} are related to the identity, permutation and Temperley-Lieb operators. Their explicit relations as given,
\begin{align}
\check{P}_{12}^{(1)}=-\tfrac{1}{2n} E_{12}, ~
\check{P}_{12}^{((2n+1)(n-1))}= \tfrac{1}{2} \left( I_{12} -  P_{12} \right) + \tfrac{1}{2n} E_{12}, 
\check{P}_{12}^{((2n+1)n)}= \tfrac{1}{2}\left( I_{12} +  P_{12} \right). \nonumber
\end{align}~

Nevertheless, instead of labeling the subspaces by the dimension of the irreducible representation for the general $Sp(2n)$, it is simpler to label it by the Dynkin labels of the representation.  Therefore, we define a shorthand notation to indicate the irreducible representations in terms of the Dynkin labels. The fundamental representations are generally denoted by $\{k\}:=[0,\dots,0,\overbrace{1}^{k-th},0,\dots,0]$. On the other hand, the one-dimensional representation is denoted by $\{0\}:=[0,\dots,0]$ and the remaining representations needed here are denoted by $\{1;k\}:=[1,0,\dots,0,\overbrace{1}^{k-th},0,\dots,0]$ for $k\neq 1$ and  $\{1;1\}=[2,0,\dots,0]$. 

The Klebsch-Gordan series $W^{\{1\}} \otimes W^{\{m\}}= \sum_{k_m} W^{\{k_m\}}$ for $m=1,\dots,n$ is represented as \cite{GROUP},
\begin{alignat}{3}
	\{1\}\otimes\{1\}&=\{0\}\oplus \{2\} \oplus\{1;1\} \nonumber\\  
	\{1\}\otimes\{2\}&=\{1\}\oplus \{3\} \oplus\{1;2\} \nonumber\\
	\vdots \nonumber\\
	\{1\}\otimes\{m\}&=\{m-1\}\oplus \{m+1\}\oplus \{1;m\} \\
	\vdots \nonumber\\
	\{1\}\otimes\{n-1\}&=\{n-2\}\oplus \{n\}\oplus \{1;n-1\} \nonumber\\	
	\{1\}\otimes\{n\}&=\{n-1\} \oplus \{1;n\}, \nonumber 
\end{alignat}
where the dimension of the fundamental representations $\{k\}$ is given by 
\eq
\dim\{k\}=\!\left(\!\begin{array}{c} \! 2n \!\\ k\end{array}\!\right)-\!\left(\!\begin{array}{c} 2n \\ \!k-2 \!\end{array}\!\right)\!. \nonumber
\en

In this notation, we have the $R$-matrices written as follows,
\bear
&&{\cal R}^{(m,1)}_{12}(\lambda)=(\lambda+\tfrac{m+1}{2})(\lambda-\tfrac{2\Delta+1-m}{2}) \check{\cal P}_{12}^{(\{m-1\})} \nonumber\\
&+& (\lambda-\tfrac{m+1}{2})(\lambda+\tfrac{2\Delta+1-m}{2}) \check{\cal P}_{12}^{(\{m+1\})}  + (\lambda+\tfrac{m+1}{2})(\lambda+\tfrac{2\Delta+1-m}{2}) \check{\cal P}_{12}^{(\{1;m\})},  
\ear
for $m=1,\dots,n-1$ and the last fusion $R$-matrix is given by,
\bear
{\cal R}^{(n,1)}_{12}(\lambda)=(\lambda-\tfrac{\Delta+2}{2}) \check{\cal P}_{12}^{(\{n-1\})} +  (\lambda+\tfrac{\Delta+2}{2}) \check{\cal P}_{12}^{(\{1;n\})},  
\ear
where $\check{\cal P}_{12}^{(\{k\})}$ are the projectors onto the irreducible space $W^{\{k\}}$ in the Klebsch-Gordan series $W^{\{m\}} \otimes W^{\{1\}}= \sum_{k_m} W^{\{k_m\}}$ for $m=1,\dots,n$. For instance, for $Sp(6)$, we have that $\check{\cal P}_{12}^{(\{1\})} :=\check{P}_{12}^{(6)}$. Besides the case $m=1$  ${\cal R}^{(1,1)}_{12}(\lambda)=R^{(2n,2n)}_{12}(\lambda)$  which is in agreement with Eq.(\ref{r2n2n}). 

This allows us to define additional transfer matrices such that the $\{m\}$ representation sits on the auxiliary space such that,
\eq
T^{(\{m\})}(\lambda)=\tr_{\cal A}{\left[{\cal T}_{\cal A}^{(m,1)}(\lambda)\right]}, \qquad {\cal T}_{\cal A}^{(m,1)}(\lambda)={\cal R}_{{\cal A}L}^{(m,1)}(\lambda){\cal R}_{{\cal A}L-1}^{(m,1)}(\lambda)\cdots {\cal R}_{{\cal A}1}^{(m,1)}(\lambda).         
\en

The fused $R$-matrices satisfies the unitarity condition,
\bear
{\cal R}_{12}^{(m,1)} (\lambda) {\cal R}_{21}^{(1,m)} (-\lambda) &=& ((\tfrac{m+1}{2})^2-\lambda^2)((\tfrac{2\Delta+1-m}{2})^2-\lambda^2) I_{12}, \\
{\cal R}_{12}^{(n,1)} (\lambda) {\cal R}_{21}^{(1,n)} (-\lambda) &=& ((\tfrac{\Delta+2}{2})^2-\lambda^2) I_{12}.
\ear

Formally, by exploiting the singular values of ${\cal R}_{12}^{(1,1)}(\lambda)$ and subsequently ${\cal R}_{12}^{(m,1)}(\lambda)$  we have that,
\bear
\check{\cal P}_{ab}^{(\{0\})} {\cal R}_{b2}^{(1,1)}(\lambda){\cal R}_{a2}^{(1,1)}(\lambda-\Delta)\check{\cal P}_{ab}^{(\{0\})} &=&(\lambda^2-1)(\lambda^2-\Delta^2)\check{\cal P}_{ab}^{(\{0\})},  \label{fusSp2n1} \nonumber\\
\check{\cal P}_{ab}^{(\{m+1\})} {\cal R}_{b2}^{(1,1)}(\lambda){\cal R}_{a2}^{(m,1)}(\lambda-\tfrac{m+1}{2})\check{\cal P}_{ab}^{(\{m+1\})}&=&(\lambda-1)(\lambda+\Delta)
{\cal R}_{12}^{(m+1,1)}(\lambda-\tfrac{m}{2}), \label{fusSp2n2} \nonumber \\
\check{\cal P}_{ab}^{(\{n\})} R_{b2}^{(1,1)}(\lambda){\cal R}_{a2}^{(n-1,1)}(\lambda-\tfrac{n}{2})\check{\cal P}_{ab}^{(\{n\})}&=&(\lambda^2-1)(\lambda+\Delta) {\cal R}_{12}^{(n,1)}(\lambda-\tfrac{n-1}{2}), \label{fusSp2n3}  \\
\check{\cal P}_{ab}^{(\{m\})} {\cal R}_{b2}^{(1,1)}(\lambda){\cal R}_{a2}^{(m+1,1)}(\lambda-\tfrac{2\Delta-m}{2})\check{\cal P}_{ab}^{(\{m\})}&=&(\lambda-1)(\lambda+\Delta) {\cal R}_{12}^{(m,1)}(\lambda-\tfrac{2\Delta-m-1}{2}), \label{fusSp2n4} \nonumber \\
\check{\cal P}_{ab}^{(\{n-1\})} {\cal R}_{b2}^{(1,1)}(\lambda){\cal R}_{a2}^{(n,1)}(\lambda-\tfrac{\Delta+2}{2})\check{\cal P}_{ab}^{(\{n-1\})}&=&(\lambda+\Delta) {\cal R}_{12}^{(n-1,1)}(\lambda-\tfrac{\Delta+1}{2}), \label{fusSp2n5} \nonumber
\ear
for $m=1,2,\dots,n-2$, which allows us to obtain the transfer matrix fusion relations (\ref{TMFISp2n3}).

\renewcommand{\baselinestretch}{1.5}

\end{document}